\documentclass{PoS}

\title{Dilaton EFT from p-regime to RMT in the $\epsilon$-regime}

\ShortTitle{Dilaton EFT from p-regime to RMT in the $\epsilon$-regime   \hskip 1.3in   {\rm Julius Kuti and}}

\author{Zoltan Fodor\\
	University of Wuppertal, Department of Physics, Wuppertal D-42097, Germany\\
	Juelich Supercomputing Center, Forschungszentrum Juelich, Juelich D-52425, Germany\\
	Eotvos University, Pazmany Peter setany 1, 1117 Budapest, Hungary\\
	\email{fodor@bodri.elte.hu}}

\author{Kieran Holland\\
	University of the Pacific, 3601 Pacific Ave, Stockton CA 95211, USA\\
	Albert Einstein Center for Fundamental Physics, Bern University, Bern, Switzerland\\
	\email{kholland@pacific.edu}}

\author{\speaker{Julius Kuti}\\
	University of California, San Diego, 9500 Gilman Drive, La Jolla, CA 92093, USA\\
	\email{jkuti@ucsd.edu}}

\author{\speaker{Chik Him Wong}\thanks{Combined contribution of both speakers}\\
	University of Wuppertal, Department of Physics, Wuppertal D-42097, Germany\\
	\email{cwong@uni-wuppertal.de}}

\abstract{
	
	New results are reported from tests of a low-energy effective field theory (EFT) that includes a  dilaton field to describe the emergent light scalar with ${ 0^{++} }$ quantum numbers in the strongly coupled near-conformal gauge theory with a massless fermion flavor doublet in the two-index symmetric (sextet) representation of the SU(3) color gauge group.  In the parlor of walking ---  based on the observed light scalar, the small $\beta$-function at strong coupling, and the large anomalous scale dimension of the chiral condensate --- 	the dilaton EFT hypothesis is introduced to test if it explains the slowly changing nearly scale invariant physics that connects the asymptotically free UV fixed point and the far-infrared scale of chiral symmetry breaking. The characteristic dilaton EFT signatures of scale symmetry breaking are probed in this report in the small Compton wavelength limit of Goldstone bosons relative to the size of the lattice volume (p-regime) and in the limit when the Goldstone wavelength exceeds the size of the volume ($\epsilon$-regime). Random matrix theory (RMT) analysis of the dilaton EFT is applied to the lowest part of the Dirac spectrum in the $\epsilon$-regime to directly test predictions for the fundamental EFT parameters. The predictions, sensitive to the choice of the dilaton potential, were limited before to the p-regime, using extrapolations from far above the chiral limit with untested uncertainties. The dilaton EFT analysis of the $\epsilon$-regime was first suggested in~\cite{Fodor:2019vmw}, with some results presented at this conference and with our continued post-conference analysis added to stimulate discussions.
}

\FullConference{The 37th Annual International Symposium on Lattice Field Theory - LATTICE2019\\
	16-22 June, 2019\\
	Wuhan, China.}

\usepackage{graphicx}
\usepackage{comment}
\usepackage{wrapfig}
\usepackage{cite}
\usepackage{subcaption}
\usepackage{amsmath}
\usepackage{setspace}
\usepackage{ragged2e}
\usepackage{grffile}
\usepackage{cleveref}

\newcommand{\be}{\begin{eqnarray}}
\newcommand{\ee}{\end{eqnarray}}

\newcommand{\cT}{{\cal T}}
\newcommand{\cD}{{\cal D}}
\newcommand{\cZ}{{\cal Z}}

\begin{document}

	\section{Introduction}
%	\vskip -0.1in
	New results are reported here from dilaton inspired hypothesis tests of a low-energy effective field theory (EFT)  to describe the emergent light scalar with ${ 0^{++} }$ quantum numbers in the strongly coupled near-conformal gauge theory with a massless fermion flavor doublet in the two-index symmetric (sextet) representation of the SU(3) color gauge group~\cite{Fodor:2017nlp,Fodor:2019vmw} (the sextet model was discussed earlier without dilaton analysis in~\cite{Hong:2004td,Dietrich:2005jn,Fodor:2012ty,Fodor:2014pqa,Fodor:2016pls}). Important and influential dilaton EFT analyses have been presented recently in the p-regime and applied to the eight-flavor model with fermions in the fundamental representation of SU(3) color~\cite{Golterman:2016lsd,Golterman:2016cdd,Appelquist:2017wcg,Appelquist:2017vyy,Golterman:2018mfm,Golterman:2018bpc,Appelquist:2019lgk}. 
		
	The dilaton EFT  hypothesis describes a slowly changing nearly scale invariant region that connects the asympotically free UV fixed point and the far-infrared scale of chiral symmetry breaking in near-conformal strongly coupled gauge theories.  In the sextet model this description is motivated by its light scalar and its small $\beta$-function, tested at strong coupling in~\cite{Fodor:2015zna,Fodor:2019ypi}, and further supported by the large anomalous scale dimension of its chiral condensate~\cite{Fodor:2016hke,Fodor:2017nlp}. The characteristic dilaton signatures of spontaneously broken scale invariance are probed here  in two distinct limits of fermion mass deformations. Both the small Compton wavelength limit of Goldstone bosons relative to the size of the lattice volume (p-regime) and the limit when the Goldstone wavelength exceeds the size of the volume ($\epsilon$-regime) are tested with two dilaton potentials of different theoretical origin. Random matrix theory (RMT) analysis of the EFT is applied to the lowest part of the Dirac spectrum in the $\epsilon$-regime.
		
	Dilaton motivated analysis of the $\epsilon$-regime has a curious recent history. The fundamental parameters of the dilaton EFT are defined in the chiral limit at vanishing fermion mass. Their safe determination and more complete tests of the EFT would require to reach down in the p-regime very close to the massless fermion limit of chiral symmetry breaking (${\chi SB}$) where Goldstone dynamics dominates, largely disentangled from the light scalar. High above the chiral limit the p-regime analysis is not a complete test of the theory. As noted in~\cite{Golterman:2018mfm,Golterman:2018bpc}, lattice ensembles in the leading order (LO), or equivalently cited as tree-level, of the dilaton EFT tests of the eight-flavor model had been fitted in~\cite{Appelquist:2017vyy,Appelquist:2017wcg,Golterman:2018mfm,Golterman:2018bpc,Golterman:2019htd} high above the chiral limit, with Goldstone bosons and the light scalar nearly degenerate without Goldstone dominance. In fact, it was argued in~\cite{Golterman:2018mfm,Golterman:2018bpc} that the estimated two orders of magnitude drop of the fermion masses to reach Goldstone dominance is out of reach for lattice investigations requiring lattice volumes in the hundreds of the lattice spacing with the pion Compton wavelength close to one hundred with insurmountable critical slowing down. Facing similar challenges in the sextet model, at Lattice 2018 we proposed a solution to this problem by switching to new lattice ensembles defined in the $\epsilon$-regime where the two orders of magnitude drop in the fermion masses becomes feasible. The EFT of the $\epsilon$-regime and the $\delta$-regime were outlined and feasibility tests were presented at Lattice 2018 to produce the required lattice ensembles~\cite{Fodor:2019vmw}. Reporting on the $\epsilon$-regime continued at this conference but incomplete tests of the dilaton EFT in the $\epsilon$-regime were not presented at the conference from insufficient statistics of our lattice ensembles.\footnote{Details of the replica method we use to describe the leading order RMT scaling laws of the dilaton EFT will be presented in a separate publication. Similar scaling laws were recently derived in~\cite{Brown:2019ipr}.} 
	The full statistics of the current $\epsilon$-regime RMT analysis is added here from increased post-conference statistics to facilitate further discussions of competing ideas in the public arena.
	
	In Section 2, in the parlor of {\em walking}, we briefly outline the most attractive walking scenario of the sextet theory from the pinch of a pair of complex conformal fixed points~\cite{Vecchi:2010jz,Gorbenko:2018ncu,Gorbenko:2018dtm}. The hypothesis of the dilaton EFT is summarized in Section 4 with new p-regime results from the sextet theory. In Section 5, tests of the dilaton EFT hypothesis in the $\epsilon$-regime are reported using RMT analysis on the lowest part of the Dirac spectrum from the current statistics of our sextet lattice ensembles. In Section 6 we close with brief conclusions. 
	
	%%%%%%%%%%%%%%%%%%%%%%%%%%%%%%%%%%%%%%%%%%%%%
	
	\vskip -0.8in	
	\section{Walking}
	
	The sextet gauge theory, investigated in this report, has an asymptotically free UV fixed point at flavor number $n_f=2$, the target of our interest. While the two-index symmetric representation of the fermions in the SU(3) color gauge group is kept fixed, the theory exhibits quantum  phase transitions at the boundaries of the conformal window (CW) as the flavor number $n_f$ is varied. 	The lower edge of the CW at $n_f^{lower}$ is expected to be slightly above $n_f=2$ signaling the onset of the conformal phase with increasing $n_f$ when the lower edge of the CW is crossed. The upper edge $n_f^{upper}$ is slightly above the weakly coupled Banks-Zaks conformal fixed point~\cite{Banks:1981nn} at $n_f=3$, with conformality and asymptotic freedom lost above $n_f^{upper}$. 
	
	The main interest for us is the strongly coupled sextet theory with $n_f=2$, just below the CW and challenging to analyze. A plausible scenario is outlined in~\cite{Kaplan:2009kr,Gies:2005as} with two fixed points colliding at $n_f^{lower}$ and moving into the complex plane as a complex conjugate pair. Just below the conformal edge, where the $n_f=2$ sextet theory sits, walking dynamics is expected to imply an energy range with approximate scale invariance. In the far-infrared, the approximately scale invariant regime would cross over into the ${\chi SB}$ phase with confinement. Using the 2d Potts model as a function of Q (the number of Potts states),  in recent influential papers Gorbenko, Rychkov, and Zan explain how walking dynamics could be understood by building the theory on the complex conjugate pair of fixed points as the realization of a complex conformal field theory (CFT)~\cite{Gorbenko:2018ncu,Gorbenko:2018dtm}. In earlier work, Vecchi captured similar ingredients of this walking scenario in 4d gauge theories just below the CW~\cite{Vecchi:2010jz}.
	
	%%%%%%%%%%%%%%%%%%%%%%%%%%%%%%%%%%%%%%%%%%%%%
	
	\vskip -0.3in
	\subsection{Pilot study of  walking in the Potts model}
	Influenced by ideas from~\cite{Vecchi:2010jz,Gorbenko:2018ncu,Gorbenko:2018dtm} 
		\begin{figure}[h!]
		\begin{center}
			\begin{tabular}{ccc}
				\includegraphics[height=3.7cm]{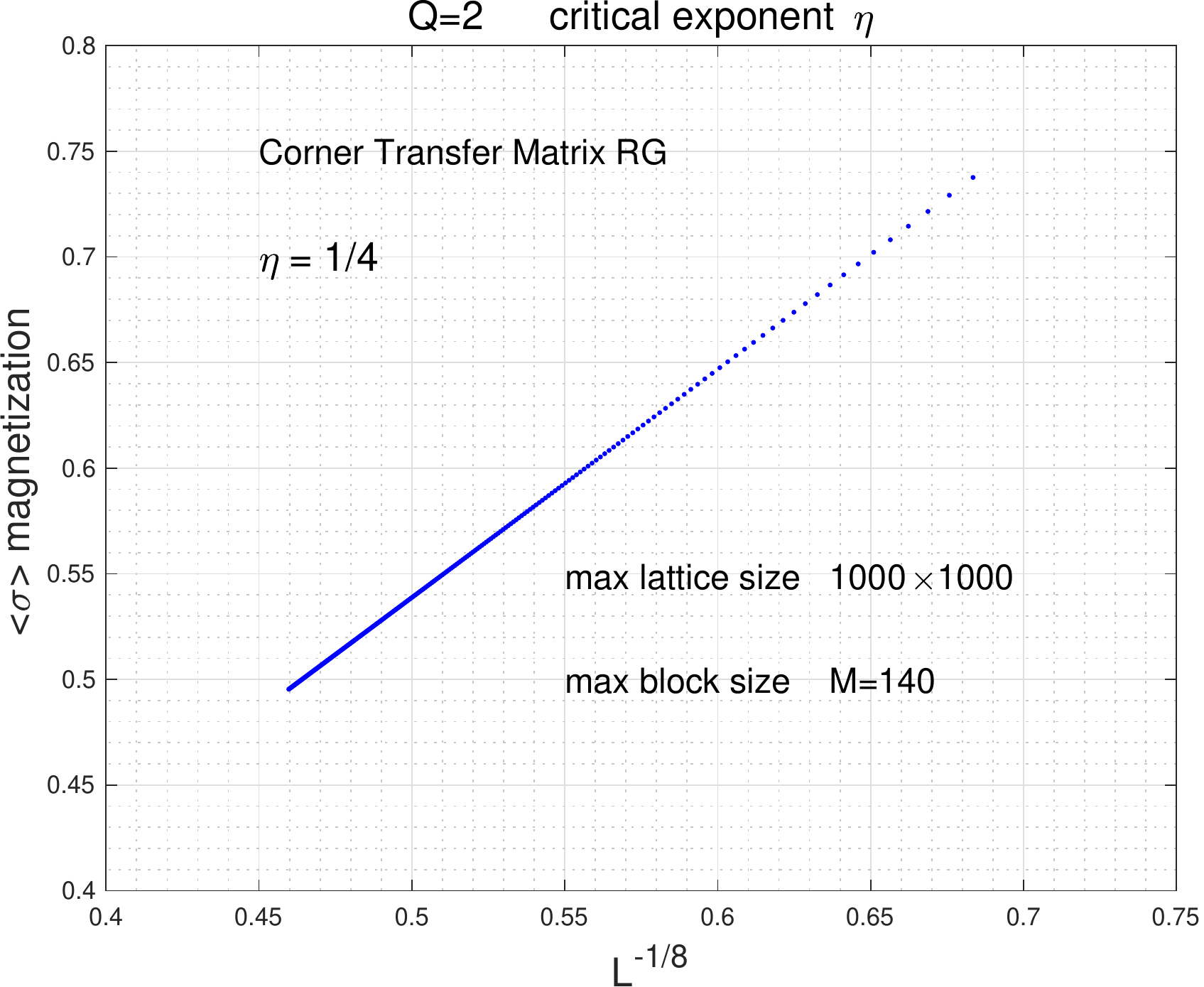}&
				\includegraphics[height=3.7cm]{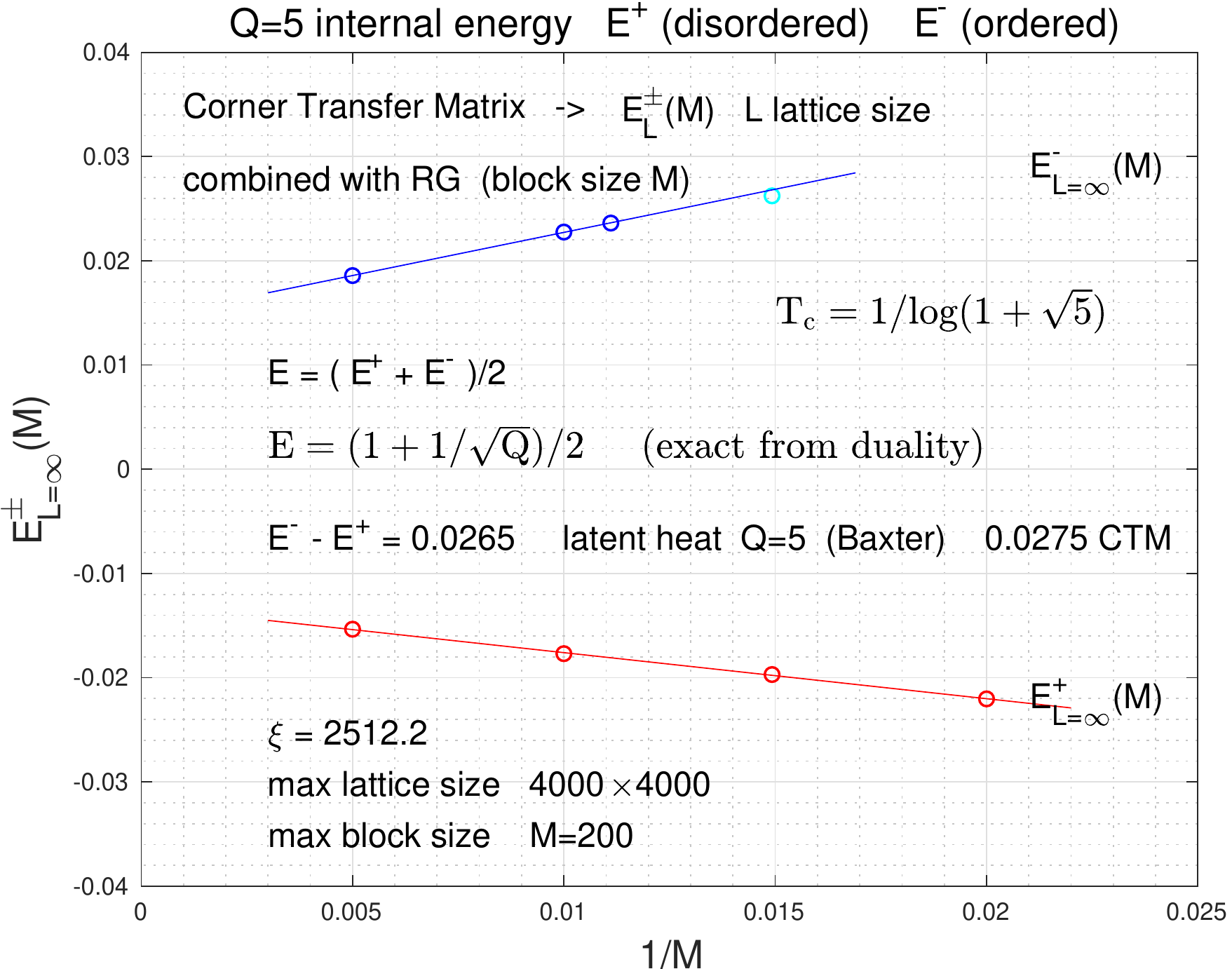}&
				\includegraphics[height=3.7cm]{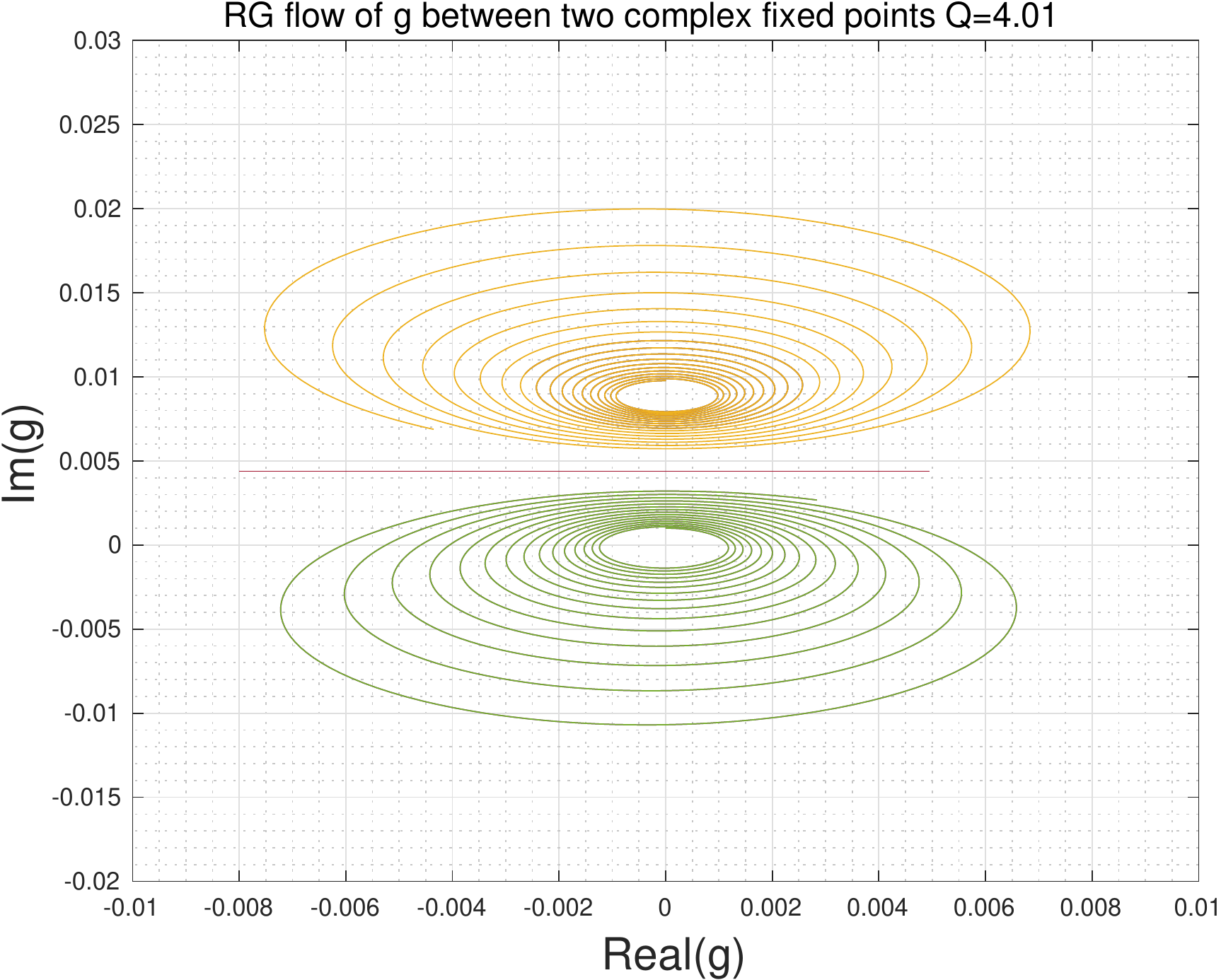}
			\end{tabular}
		\end{center}	
		\caption{\footnotesize  The critical Potts exponent $\eta$ is shown on the left panel from CTM calculations at Q=2 from finite size scaling analysis of the RG block size $L$ as it is varied inside the 1000x1000 CTM lattice~\cite{Baxter:1978,Nishino:1996}.
			The central panel shows the latent heat determination at Q = 5 as the block size M of the large 4000x4000 CTM is varied in the plotted range 
			with similar earlier result from the authors of~\cite{Nishino:1996}.  The right panel shows the unwinding of 
	    	the flow driven by near-marginal operator when the complex conjugate fixed points are perturbed at Q=4.01. The straight line in-between is the walking RG flow of the theory for Q > 4~\cite{Gorbenko:2018ncu,Gorbenko:2018dtm} .}
		\label{fig:potts}
	\end{figure}
    we started our own investigation of walking behavior in the 2d Potts model with the first results presented at the conference. The main goal is to support with concrete lattice realization the underlying continuum theory of~\cite{Gorbenko:2018ncu,Gorbenko:2018dtm} for continuous variations of Q, the number of Potts states. 
    The model has a weakly first-order phase transition and walking behavior in the range 4 < Q < 10. 	For Q < 4,  the Potts model has a critical and a tricritical point, and the corresponding partition functions on the torus are known and the full complex CFT of the continuum model can be obtained by analytically continuing these partition functions to Q > 4~\cite{Gorbenko:2018ncu,Gorbenko:2018dtm}.  
 	Walking dynamics can be described then in Potts model setting as the complex CFT perturbed by a nearly marginal operator, making predictions for observable quantities. Since the walking regime is only approximately scale invariant, 2pt functions exhibit small deviations from power laws, computed in perturbation theory. One of our primary goals is to test the predicted drifting scaling dimensions~\cite{Gorbenko:2018ncu,Gorbenko:2018dtm}  in lattice simulations. We adopted the Corner Transfer Matrix method~\cite{Baxter:1978,Nishino:1996} as one of our primary lattice tools with some  pilot analysis shown in Fig.~\ref{fig:potts} to reproduce earlier results in the Potts model and to apply it to the correlation functions with drifting exponents. 
    
  \subsection{Four-dimensional gauge theories}
  
  Generalizing walking dynamics from the Potts model, conformality could be lost in near-conformal 4d gauge theories below the CW when two  fixed points collide and move into the complex plane. Walking dynamics  would be controlled by scaling dimensions of operators in the (non-unitary) CFTs living at the pair of complex conjugate fixed points \cite{Gorbenko:2018ncu, Gorbenko:2018dtm}, with similar ideas in~\cite{Vecchi:2010jz}. The RG flow, defined by the coupling of the operator perturbing the complex fixed point, would occur between the two  complex fixed points pinching the real axis where the physical coupling evolves. The walking regime ends when the renormalization group flow of the coupling leads to the chiral symmetry breaking phase in the far infrared. A light scalar particle would appear in the spectrum, perhaps with dilaton-like signatures, but not necessarily light parametrically in comparison with other excitations. A new idea for the near-conformal dilaton scenario was presented recently in the high charge limit of the 3d U(1) model with promise for 4d gauge theory realizations~\cite{Orlando:2019skh}.
  
%%%%%%%%%%%%%%%%%%%%%%%%%%%%%%%%%%%%%%%%%%%%%	
	
	\section{New  dilaton analysis with extended data sets}
	
	Input to our earlier dilaton EFT analysis was described in Section 3 of~\cite{Fodor:2019vmw}  where $M_\pi, F_\pi, M_d$ data sets of the $n_f=2$ sextet theory were used in the $m=0.0015-0.0040$ fermion mass range, with lattices volumes from $32^3\times 64$ to $64^3\times 96$. We used infinite volume extrapolations of the  $M_\pi,F_\pi$ data sets at each of the  four input fermion masses and applied them to the dilaton analysis at fixed bare gauge coupling $\beta = 6/g^2$ of 3.20. The finite size scaling analysis of the $M_\pi, F_\pi$ data set was presented earlier in~\cite{Fodor:2017nlp}. The $M_d$ input for the light scalar was always taken from the largest volume of the lattice ensembles at each input fermion mass. The constraint equations from the dilaton EFT were evaluated approximately in the statistical analysis by fits to the constraints. 
	
   We implement here the exact implicit maximum likelihood (IML) analysis to machine precision in the evaluation of constraints, derived  in the p-regime  from  dilaton scaling relations of the EFT~\cite{Appelquist:2017wcg,Appelquist:2017vyy,Golterman:2016lsd}. The exact IML procedure gives increased confidence in the determination of the five fundamental parameters 
   of the dilaton EFT for the two investigated choices of the dilaton potential. 
  % Results from the $\beta=3.20$ p-regime analysis of the $V_\sigma$ dilaton potential are essentially unchanged, but significant changes are reported here in the  $\beta=3.20$ p-regime analysis of the parameters from the hypothesis of the $V_d$ potential using the exact IML method. 
   For fine-grained lattice spacing, in the current work we added the $\beta=3.25$ data set to the p-regime analysis and extended the dilaton EFT analysis at both lattice spacings to the RMT Dirac spectra of the $\epsilon$-regime where the Compton wavelength of the Goldstones (named as pions below) exceeds the size of the  large $64^4$ lattice volume with the RMT analysis as an important cross-check on the LO p-regime fits of the fundamental parameters and the accuracy level of the LO approximation. 
   \vskip -1.5in
   
   %%%%%%%%%%%%%%%%%%%%%%%%%%%%%%%%%%%%%%%%%%%%%
	
	\section{Dilaton EFT analysis in the p-regime}
		\label{section:dilaton}
	\vskip -0.1in
	\noindent{\bf 4.1 ~~Dilaton EFT and its IML constraints:}
	In the currently accessible range  of fermion mass deformations in the sextet theory, the mass of the light scalar, with resonances far separated, is tracking closely the Goldstone boson (pion) triplet from spontaneous chiral symmetry breaking of the  ${ SU(2)\times SU(2)}$ flavor group. The minimal modification of the chiral Lagrangian with dilaton couplings leads to the recently investigated low-energy 
	EFT~\cite{Golterman:2016cdd,Golterman:2016lsd,Appelquist:2017wcg,Appelquist:2017vyy} to describe the light 
	scalar particle with $0^{++}$  quantum numbers, coupled to pion dynamics as a dilaton from broken scale invariance,
	\vskip -0.1in
	\begin{equation}
	{\cal L} = \frac{1}{2}\partial_{\mu} \chi \partial_{\mu} \chi\,-\,V(\chi) + \frac{f^2_\pi}{4}\big(\frac{\chi}{f_d}\big)^2 
	~{\rm tr}\big[\partial_{\mu}\Sigma~\partial_{\mu}\Sigma^\dagger\big] +
	\frac{m^2_\pi f^2_\pi}{4}\big(\frac{\chi}{f_d}\big)^y ~ {\rm tr}\big[\Sigma + \Sigma^\dagger\big],\label{eq:EFT}
	\end{equation}
	\vskip -0.1in
		\noindent where the notation $\chi = f_d\cdot e^{\sigma/f_d}$  is introduced for the  connection between the dilaton field $\sigma(x)$ and the compensator field $\chi(x)$ which 	transforms as $\chi(x) \rightarrow \chi'(x') =e^\omega \chi(x)$ under the shift $\sigma(x) \rightarrow \sigma'(x')=\sigma(x) + \omega\cdot f_d$ for scale transformations $x_\mu \rightarrow x'_\mu = e^{-\omega}x_\mu$. The notation $f_d$ designates the minimum of the dilaton potential $V(\chi)$ in the chiral limit of vanishing fermion masses. 
		The Goldstone pions in Eq.(\ref{eq:EFT}) are described by the unitary matrix field $\Sigma~=~{\rm exp}[2i\pi/f_\pi]$ 
		where the pion field is represented as $\pi=\Sigma_a\pi^aT^a$ with $n_f^2-1$ generators of the $SU(n_f)$ flavor group. 
		
		We keep the same notation as in~\cite{Appelquist:2017wcg,Appelquist:2017vyy}  for the convenience of easy 
		%for the parameters in Eqs.(\ref{eq:EFT},\ref{eq:Vd},\ref{eq:Vsigma}) for the convenience of easy 
		comparison between the $n_f=8$ fundamental rep and the $n_f=2$ sextet rep of separate studies. In this notation, the tree level pion mass would be $m^2_\pi = 2B_\pi m$ close to the chiral limit,  
		with the dilaton decoupled from pion dynamics.
		The pion decay constant $f_\pi$ is defined in the chiral limit. The tree-level dilaton mass in the chiral limit
		of vanishing fermion mass is designated as $m_d$ and it is
		defined by the second derivative of the tree-level dilaton potential at its $\chi = f_d$ minimum as $V''(\chi=f_d)=m_d^2$. The dilaton mass
		at finite fermion mass deformations is designated by $M_d$.

	The first application of Eq.~(\ref{eq:EFT}) to the $n_f=8$ model was reported in~\cite{Appelquist:2017wcg,Appelquist:2017vyy,Golterman:2018mfm,Golterman:2018bpc}. After our earlier  
	analysis of the sextet model in~\cite{Fodor:2019vmw,Fodor:2017nlp} we report here a broader scope of dilaton EFT tests, as indicated above for the sextet theory. 
	In Eq.~(\ref{eq:EFT}) of the EFT two different forms of the dilaton potential were chosen for analysis before,
	\vskip -0.25in
	\begin{subequations}
		\begin{align}
		V(\chi) \rightarrow V_d(\chi)  &= \frac{m^2_d}{16f^2_d}~\chi^4\big(4~{\rm ln} \frac{\chi}{f_d} - 1\big),  \label{eq:Vd} \\
		V(\chi) \rightarrow V_\sigma(\chi) &= \frac{m^2_d}{8f^2_d}~(\chi^2-f_d^2)^2.
		\label{eq:Vsigma}
		\end{align}
	\end{subequations}
	\vskip -0.1in
	Recent theoretical motivation of Eq.~(\ref{eq:Vd}) originates from~\cite{Golterman:2016lsd}, based on a parametric expansion of $V(\chi)$ as the CW is approached in the Veneziano limit of fermions with large flavor number in the fundamental representation.  The dilaton potential $V_\sigma$ in Eq.~(\ref{eq:Vsigma}) could originate from operators with small explicit breaking of scale symmetry~\cite{Goldberger:2008zz}. We do not discuss further the two choices for the dilaton potential in the sextet theory. The primary focus is to explore how dilaton EFT tests in the p-regime and the RMT analysis of the $\epsilon$-regime can probe different choices for the dilaton potential.  
	
	The Lagrangian of the dilaton EFT in Eq.~(\ref{eq:EFT}) has a long history which includes~\cite{Migdal:1982jp,Ellis:1984jv,Bardeen:1985sm,Leung:1989hw,Donoghue:1991qv,
		Goldberger:2008zz,Chacko:2012sy,Vecchi:2010gj,Matsuzaki:2013eva,Aoki:2016wnc,Hansen:2016fri,Cata:2018wzl} with further references. 
	The scale-dependent anomalous dimension $\gamma$ of the chiral condensate, with $y=3-\gamma$  in Eq.~(\ref{eq:EFT}),  features prominently in the dilaton EFT.  This raises questions about the scale-dependence of the drifting exponent $y$ in walking theories which will depend on the precise theoretical framework leading to Eq.~(\ref{eq:EFT}) of the EFT. Although this question is not addressed here, in the sextet model we have important information on the scale-dependent $\gamma$~\cite{Fodor:2016hke,Fodor:2017nlp} which can be compared with the results emerging from the analysis of Eqs.~(\ref{eq:EFT},\ref{eq:Vd},\ref{eq:Vsigma}) at fixed $y$.

	In the LO application of the dilaton EFT, implicit Maximum Likelihood (IML) procedure is used to target the five fundamental  parameters  $f_\pi, B_\pi, y, m_d,f_d$  which are defined in Eq.~(\ref{eq:EFT}).
	For the choice of the dilaton potential $V_d(\chi)$  three 	non-linear constraints at each input fermion mass $m$ are derived~\cite{Appelquist:2017wcg,Appelquist:2017vyy} and fed into the IML procedure,
	\begin{align} 
	&M_\pi^2\cdot F_\pi^{2-y} - 2B_\pi\cdot f_\pi^{(2-y)}\cdot m=0, \label{eq:Cd}\\
	&F_\pi^{(4-y)}\cdot{\rm log}(F_\pi/f_\pi) - y\cdot n_ff_\pi^{(6-y)}B_\pi\cdot m/m_d^2f_d^2=0, \label{eq:Ad}\\
	%
	%&M_\pi^2/F_\pi^2  - (2m_d^2f_d^2/yn_ff_\pi^4)\cdot{\rm log}(F_\pi/f_\pi) =0, \label{eq:Dd}\\
	%
	&(F_\pi^2/M_\pi^2)\cdot(3{\rm log}(F_\pi/f_\pi)+1) - (M_d^2/m_d^2)\cdot (f_\pi^2/M_\pi^2) - y(y-1) n_ff_\pi^4/2m_d^2f_d^2 =0. \label{eq:Md}
	\end{align} 
	The general scaling relation of Eq.~(\ref{eq:Cd}) is independent from the choice of the dilaton potential~\cite{Golterman:2016cdd,Appelquist:2017wcg}. 
	The dilaton potential  $V_d$  leads to two added non-linear conditions, with Eq.~(\ref{eq:Ad}) set by\linebreak
	$V'_d(\chi=F_d)$, and Eq.~(\ref{eq:Md}) 
	set by $V''_d(\chi=F_d)$, as  in~\cite{Appelquist:2017wcg}.
	With unchanged scaling relation from Eq.~(\ref{eq:Cd}), two alternative equations are 
	derived from $V'_\sigma(\chi=F_d)$ and $V''_\sigma(\chi=F_d)$,
	\vskip -0.3in
	\begin{align} 
	%M_\pi^2\cdot F_\pi^{2-y}  &= C\cdot m,  & C&=2B_\pi\cdot f_\pi^{(2-y)},\label{eq:C1}\\
	%
	&F_\pi^{(4-y)}\cdot(1-f_\pi^2/F_\pi^2) -  2y\cdot n_ff_\pi^{(6-y)}B_\pi/m_d^2f_d^2\cdot m = 0, \label{eq:A1}\\
	%
	%&F_\pi^2/M_\pi^2 - f_\pi^2/M_\pi^2 - y\cdot n_ff_\pi^4/m_d^2f_d^2 = 0, \label{eq:D1}\\
	%
	&3F_\pi^2/M_\pi^2-f_\pi^2/M_\pi^2 - 2M_d^2/m_d^2\cdot f_\pi^2/M_\pi^2 - y(y-1) n_ff_\pi^4/m_d^2f_d^2 =0.\label{eq:Md1}
	\end{align} 
	%	
	%\subsection{IML analysis of two dilaton potentials  in the p-regime at two gauge couplings}
	\label{section:beta320}
	We will apply IML analysis to the p-regime constraints of Eqs.~(\ref{eq:Cd}-\ref{eq:Md1}) for the two dilaton potentials with two gauge couplings representing the variation of the lattice spacing.
	
	%%%%%%%%%%%%%%%%%%%%%%%%%%%%%%%%%%%%%%%%%%%%%
	
	\vskip 0.1in
	\noindent{\bf 4.2 ~~ Gauge coupling $\mathbf {6/g^2 = 3.20:}$}
	
\begin{figure}[h!]
		\centering
		\parbox{.53\textwidth}{
			\begin{subfigure}{.49\linewidth}
				\caption*{\footnotesize posterior $f_\pi$:}
				\includegraphics[width=\textwidth]{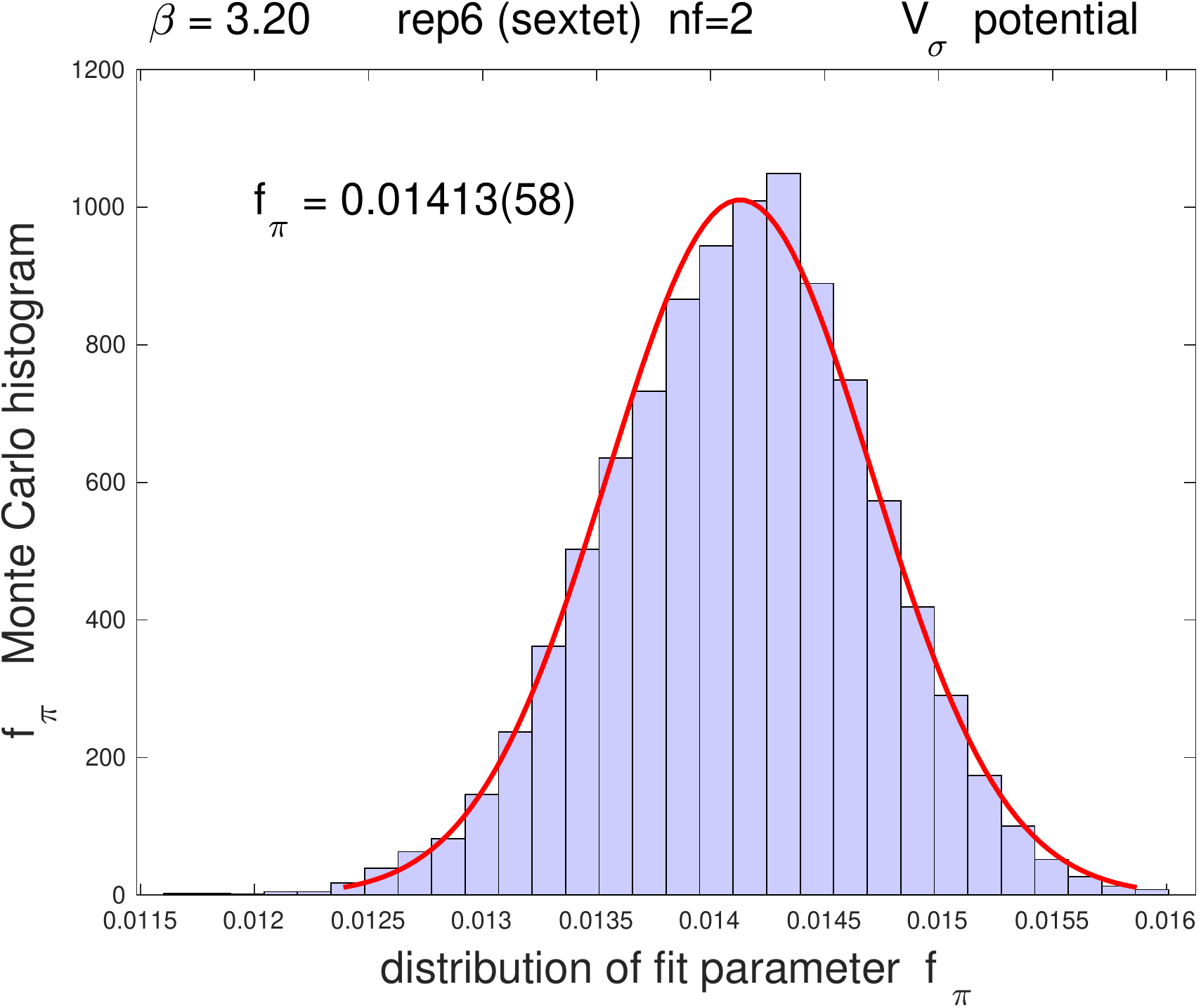}
			\end{subfigure}
			\begin{subfigure}{.49\linewidth}
				\caption*{\footnotesize posterior $\Sigma (0)=B_\pi\cdot f_\pi^2$:}
				\includegraphics[width=\textwidth]{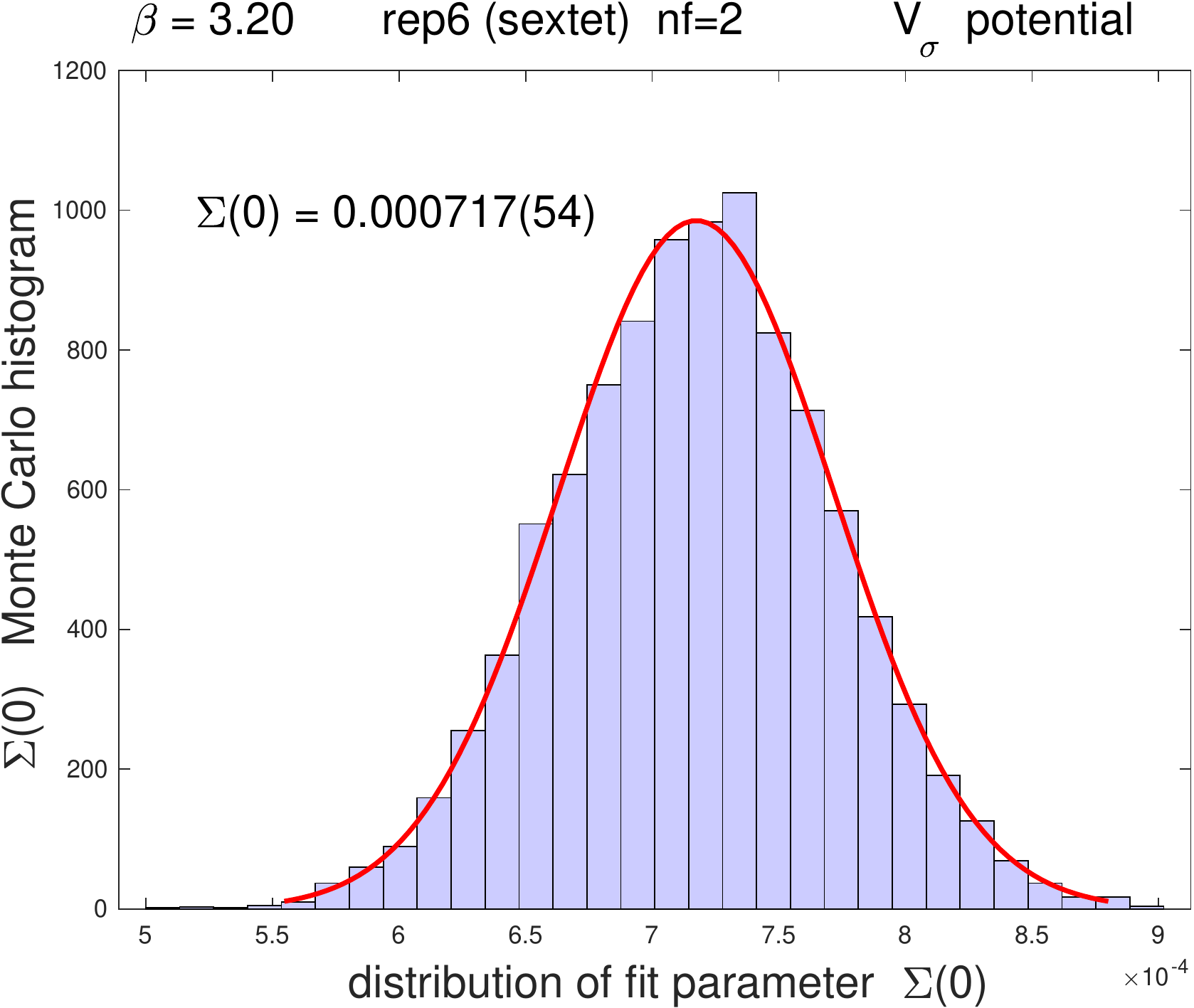}
				%			\caption{\footnotesize fit result for $B_\pi$}
			\end{subfigure}\\
			\begin{subfigure}{.49\linewidth}
				\caption*{\footnotesize posterior $\gamma$:}
				\includegraphics[width=\textwidth]{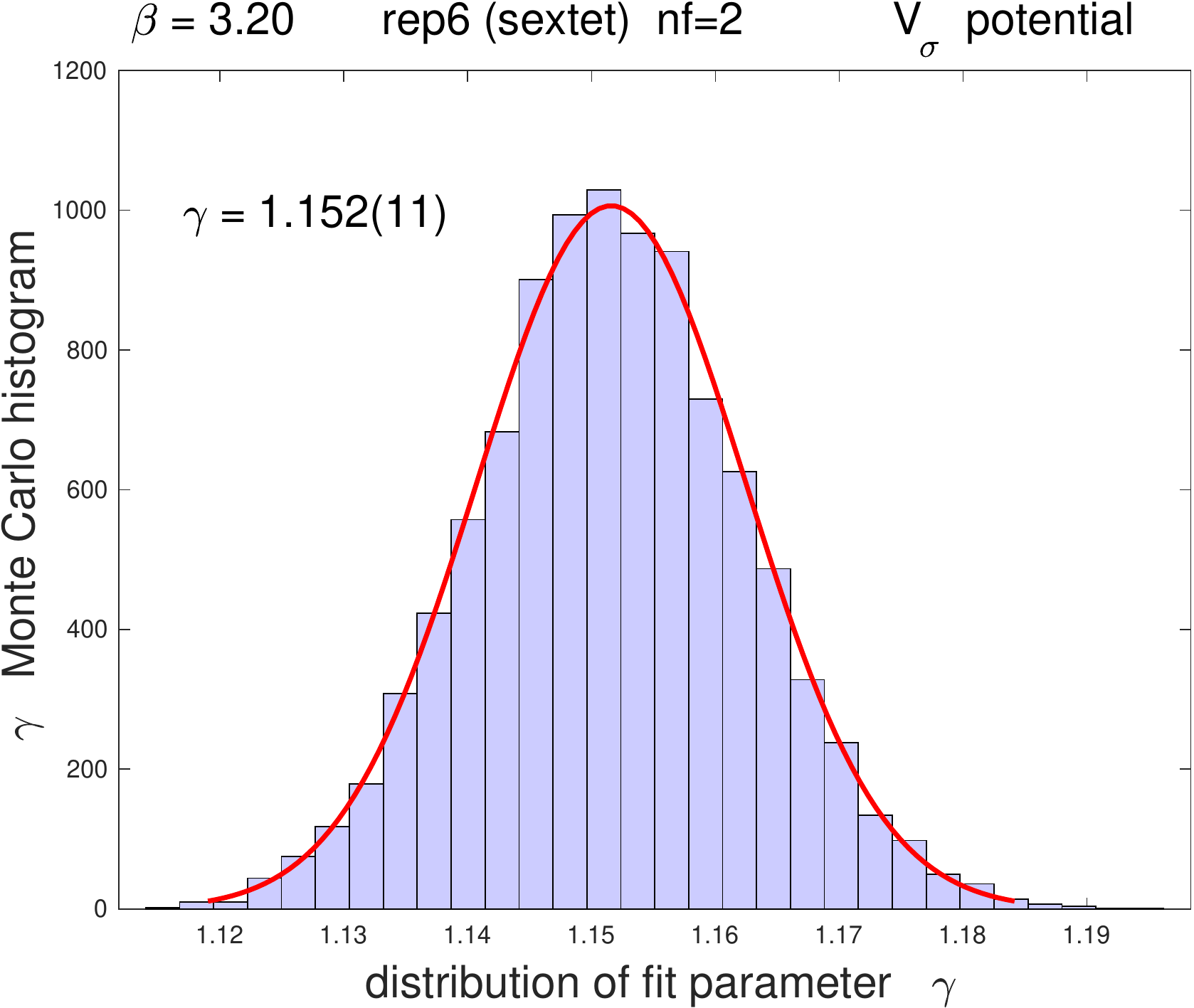}
			\end{subfigure}
			\begin{subfigure}{.49\linewidth}
				\caption*{\footnotesize posterior $m_d/f_\pi$:}
				\includegraphics[width=\textwidth]{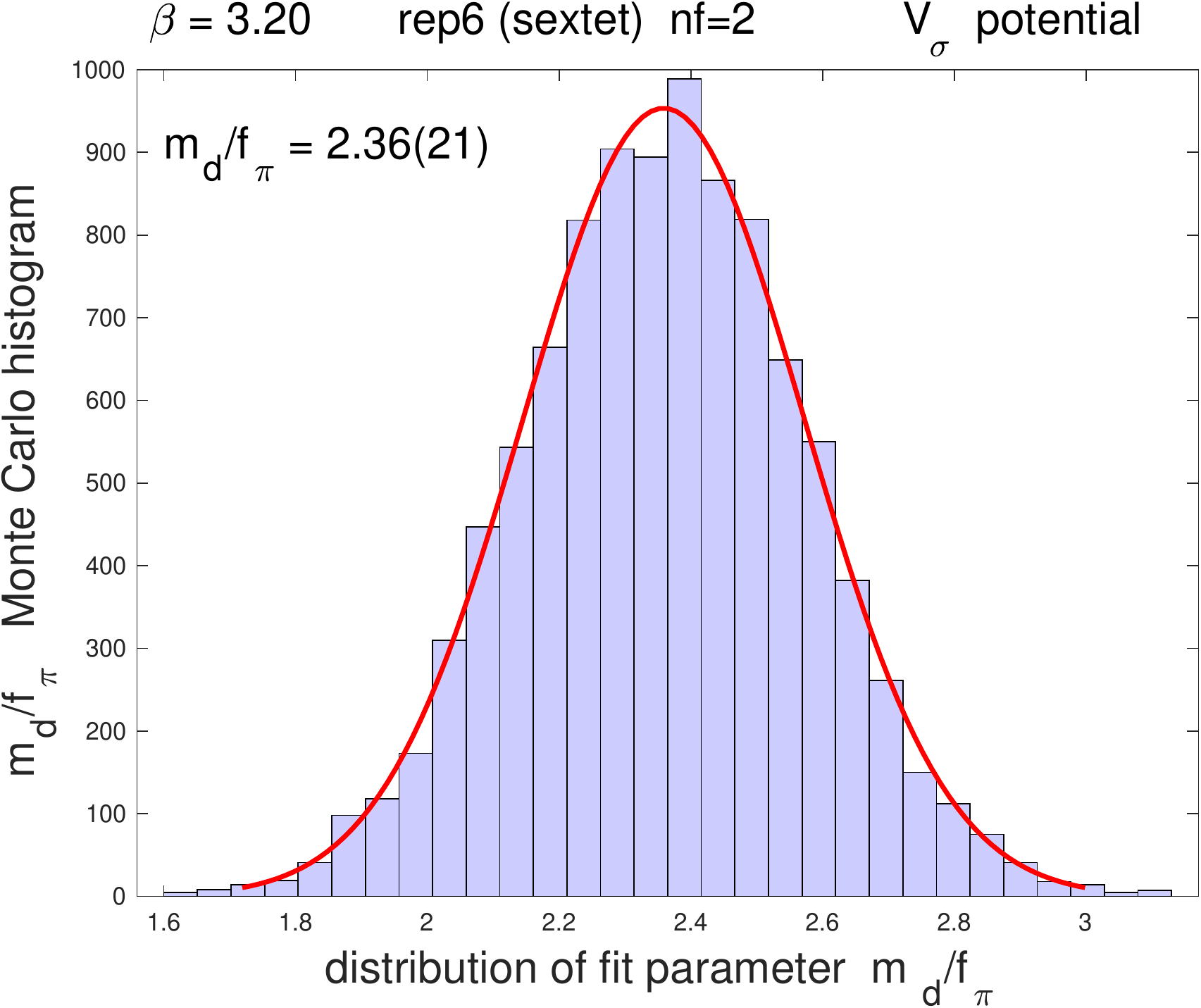}
			\end{subfigure}
		}
		\begin{subfigure}{.44\textwidth}
			\includegraphics[width=\textwidth]{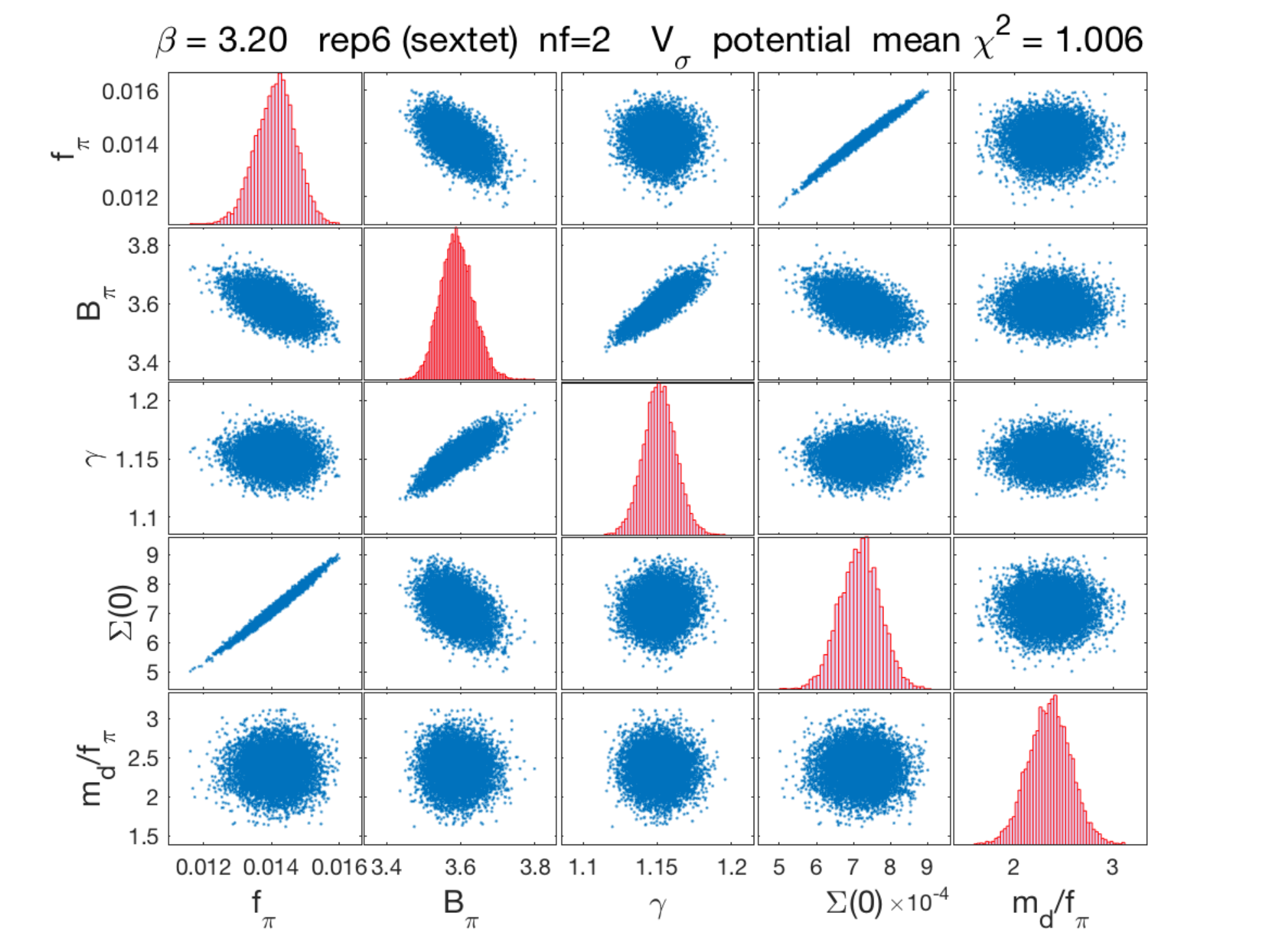}
			\caption*{\footnotesize Matrix plot for some combinations of the five independent parameters of Eq.~(\ref{eq:EFT}) 
				with posterior histograms of the Bayesian IML analysis in the diagonal for the $V_\sigma$ potential.
				Correlations are shown as  off-diagonal scatter plots. The 
				chiral condensate $\Sigma$ is shown, as obtained from the posterior distribution of the GMOR relation $\Sigma(0) = B_\pi\cdot f_\pi^2$  in the chiral limit. 
			}
		\end{subfigure}\hfill
		\caption{\footnotesize  The five independent parameters of Eq.~(\ref{eq:EFT}) were determined, based on posterior 
			distributions  from the  exact IML based analysis of the $V_\sigma$ potential, illustrated on the four left panels.  Average values and
			1$\sigma$ percentile errors are consistent with normal distributions, displayed by solid red line fits. The histograms 
			with average values $B_\pi = 3.590(45)$, and $f_d/f_\pi = 3.27(30)$ are not shown.
			The $\chi^2=1.006$ average of the posterior $\chi^2$ distribution indicates a consistent fitting procedure for the $V_\sigma$ dilaton potential.
		}
		\label{fig:rep6Vsigma320}
	\end{figure}

\begin{figure}[h!]
	\centering
	\parbox{.53\textwidth}{
		\begin{subfigure}{.49\linewidth}
			\caption*{\footnotesize posterior $f_\pi$:}
			\includegraphics[width=\textwidth]{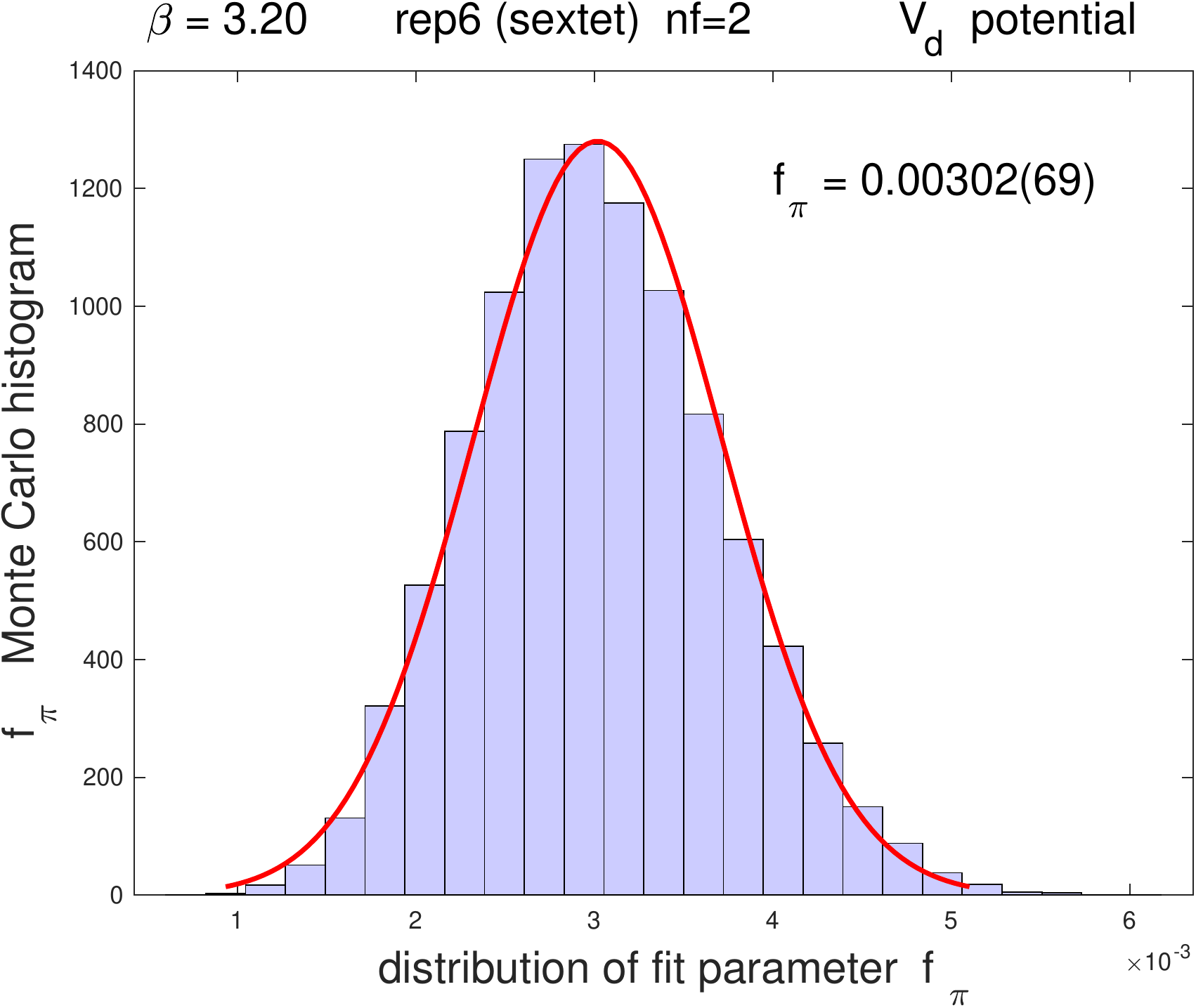}
		\end{subfigure}
		\begin{subfigure}{.49\linewidth}
			\caption*{\footnotesize posterior  $\Sigma(0) = B_\pi\cdot f_\pi^2$:}
			\includegraphics[width=\textwidth]{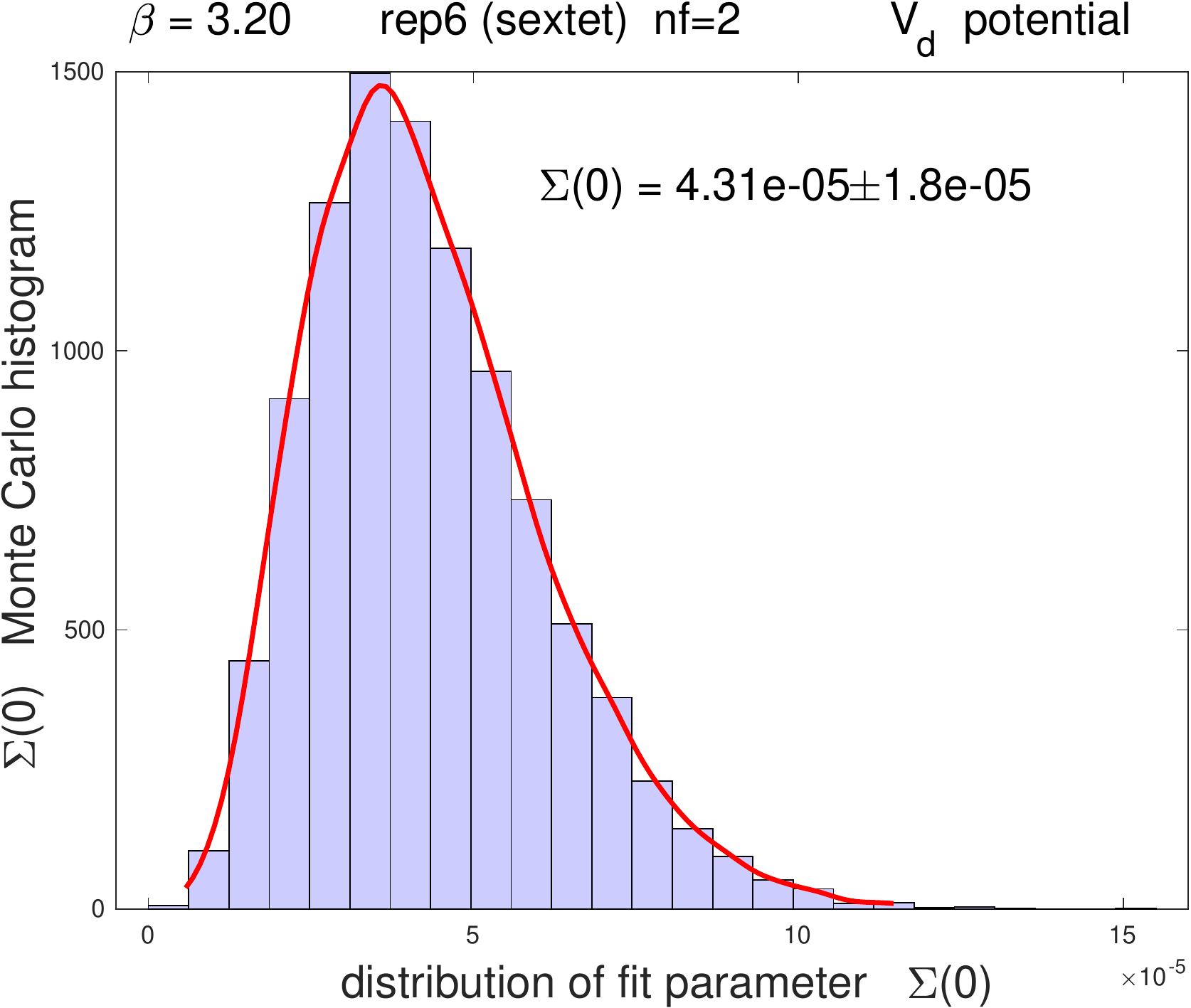}
		\end{subfigure}\\
		\begin{subfigure}{.49\linewidth}
			\caption*{\footnotesize posterior $\gamma$ :}
			\includegraphics[width=\textwidth]{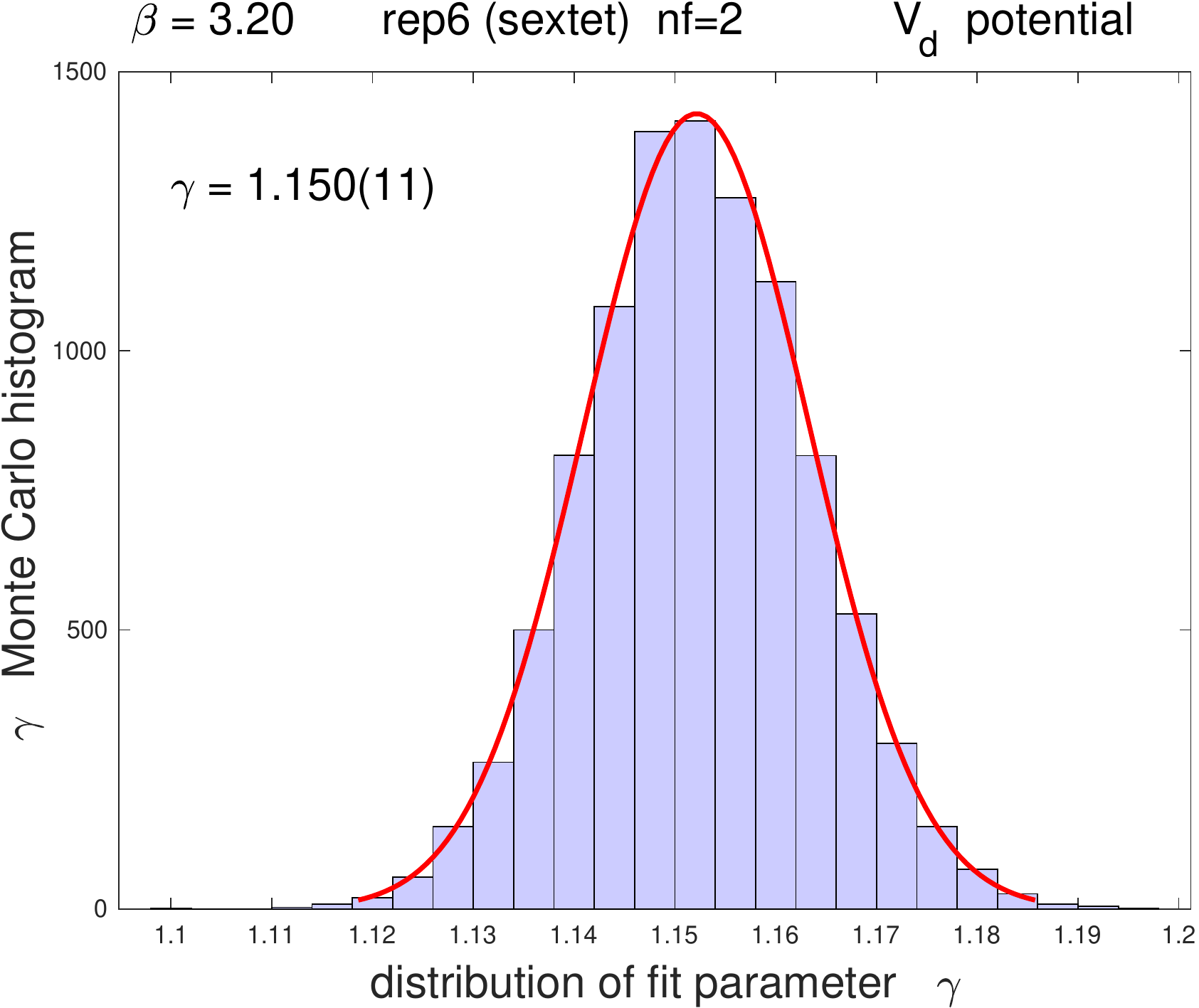}
		\end{subfigure}
		\begin{subfigure}{.49\linewidth}
			\caption*{\footnotesize posterior $m_d/f_\pi$ :}
			\includegraphics[width=\textwidth]{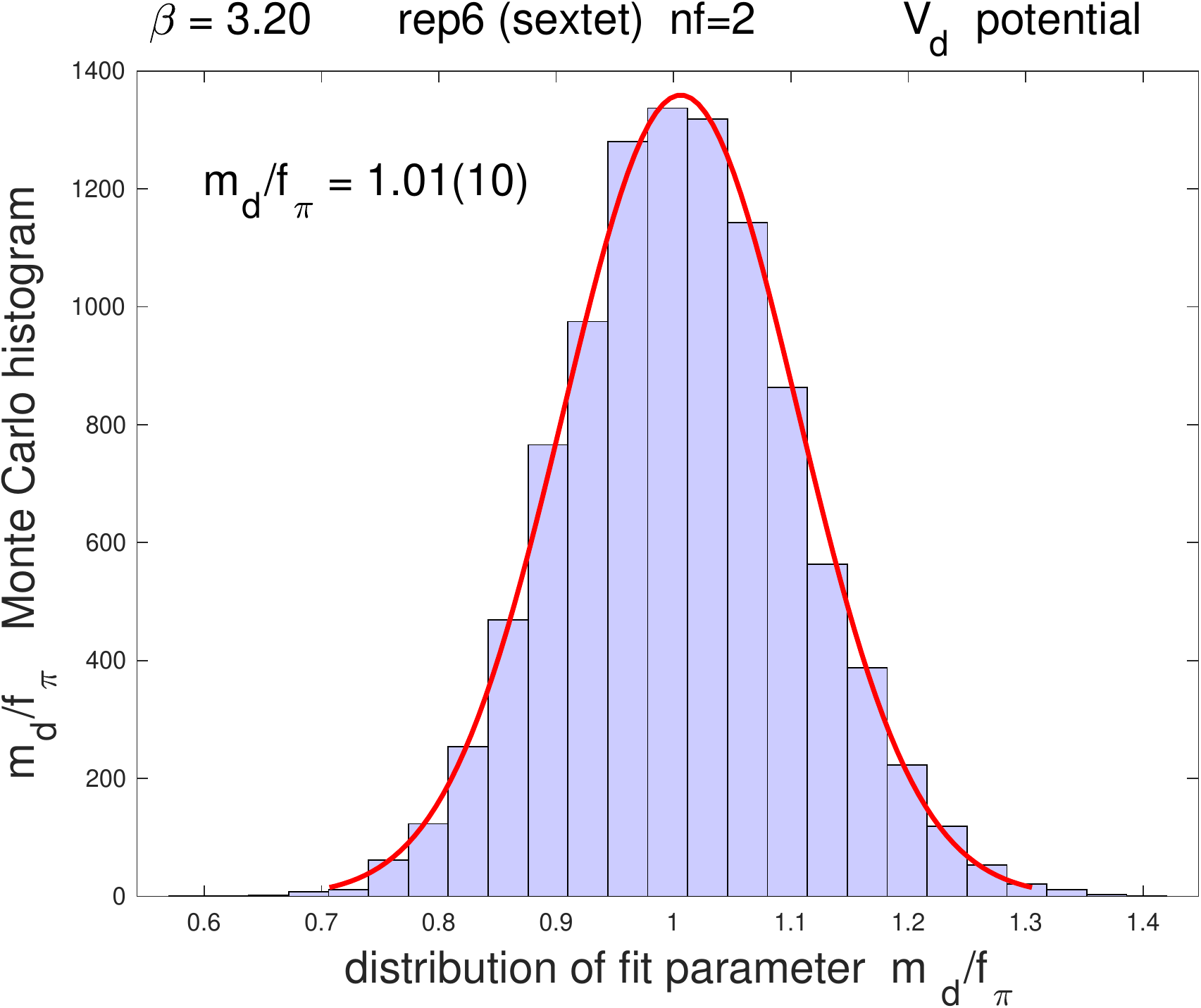}
		\end{subfigure}
	}
	\begin{subfigure}{.44\textwidth}
		\includegraphics[width=\textwidth]{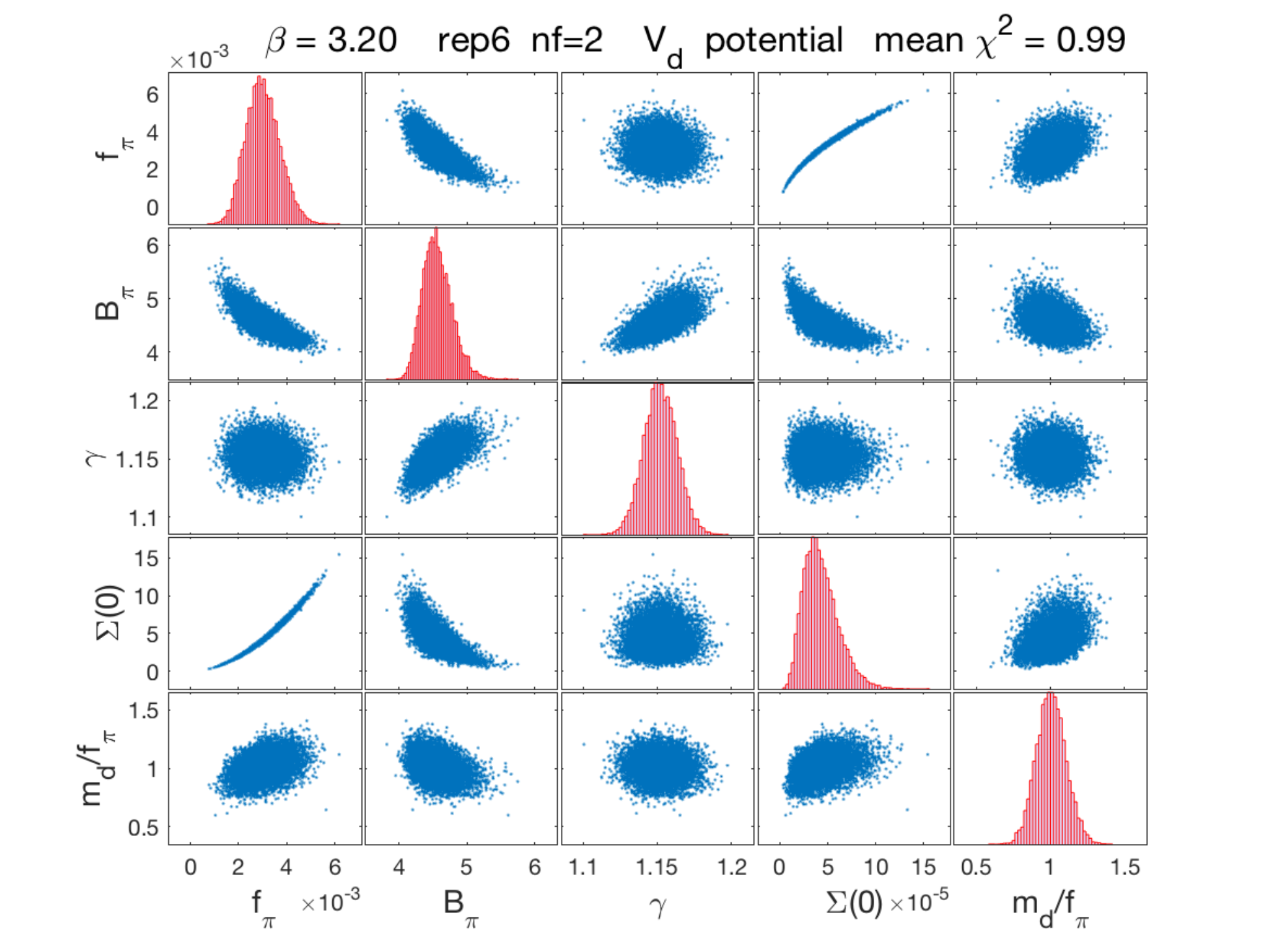}
		\caption*{\footnotesize  Matrix plot for some combinations of the five independent parameters of Eq.~(\ref{eq:EFT}) 
			with posterior histograms of the Bayesian IML analysis in the diagonal for the $V_d$ potential.
			Correlations are shown as  off-diagonal scatter plots. The 
			chiral condensate $\Sigma$ is shown, as obtained from the posterior distribution of the GMOR relation $\Sigma(0) = B_\pi\cdot f_\pi^2$  in the chiral limit. }
	\end{subfigure}\hfill
%	\vskip -0.1in
	\caption{\label{fig:rep6Vd320}\footnotesize  The five independent parameters of Eq.~(\ref{eq:EFT}) were determined, based on posterior 
		distributions  from the  exact IML based analysis of the $V_d$ potential and illustrated on the four left panels.  Average values and
		1$\sigma$ percentile errors are consistent with normal distributions, displayed by solid red line fits. The histograms 
		with average values $B_\pi = 4.57(22)$, and $f_d/f_\pi = 3.15(29)$ are not shown.
		The $\chi^2=0.99$ average of the posterior $\chi^2$ distribution would indicates consistent fitting 
		but the extreme low value of $f_\pi$ casts doubts on the LO  $V_d$ analysis of Eq.~(\ref{eq:EFT}). 
	}
\end{figure}

    	Based on the EFT Lagrangian in Eq.~(\ref{eq:EFT}), the fitted posterior statistical distributions of the physical parameters $f_\pi, B_\pi, y, m_d/f_\pi,f_d/f_\pi$ and their correlations are shown in Fig.~\ref{fig:rep6Vsigma320} for the $V_\sigma$ potential of Eq.~(\ref{eq:Vsigma}) and in Fig.~\ref{fig:rep6Vd320} for the $V_d$ potential of Eq.~(\ref{eq:Vd}). The analysis of our previous report in~ \cite{Fodor:2019vmw}, with only approximately fitted constraints, is made exact here  by imposing the constraints of Eqs.~(\ref{eq:Cd}-\ref{eq:Md1}) to machine accuracy in the IML procedure. This led to some modifications of the earlier results. The input from our lattice ensembles into the IML procedure is the same as described in~\cite{Fodor:2019vmw}. 
	%
	%
%	\newpage
    %	
	\vskip 0.1in
	For each feed from the  $M_\pi(m)$, $F_\pi(m), M_d(m)$ statistical distributions, described in~\cite{Fodor:2019vmw}, 12 input data were selected for input using fermion masses of $m=0.015/0.020/0.030/0.040$ for the exact non-linear IML minimization of the five  parameters at optimal values of the maximum likelihood functions $M^{ML}_\pi(m),F^{ML}_\pi(m), M^{ML}_d(m)$, implicitly defined by respective constraints of Eqs.~(\ref{eq:Cd}-\ref{eq:Md1}) for the two dilaton potentials. The procedure was identical at gauge coupling $6/g^2=3.25$.
		
	\label{section:beta325}
	\noindent{\bf 4.3 ~~ Gauge coupling $\mathbf {6/g^2 = 3.25:}$}
    \vskip -0.1in
	\begin{figure}[h!]
		\centering
		\parbox{.53\textwidth}{
			\begin{subfigure}{.49\linewidth}
				\caption*{\footnotesize posterior $f_\pi$:}
				\includegraphics[width=\textwidth]{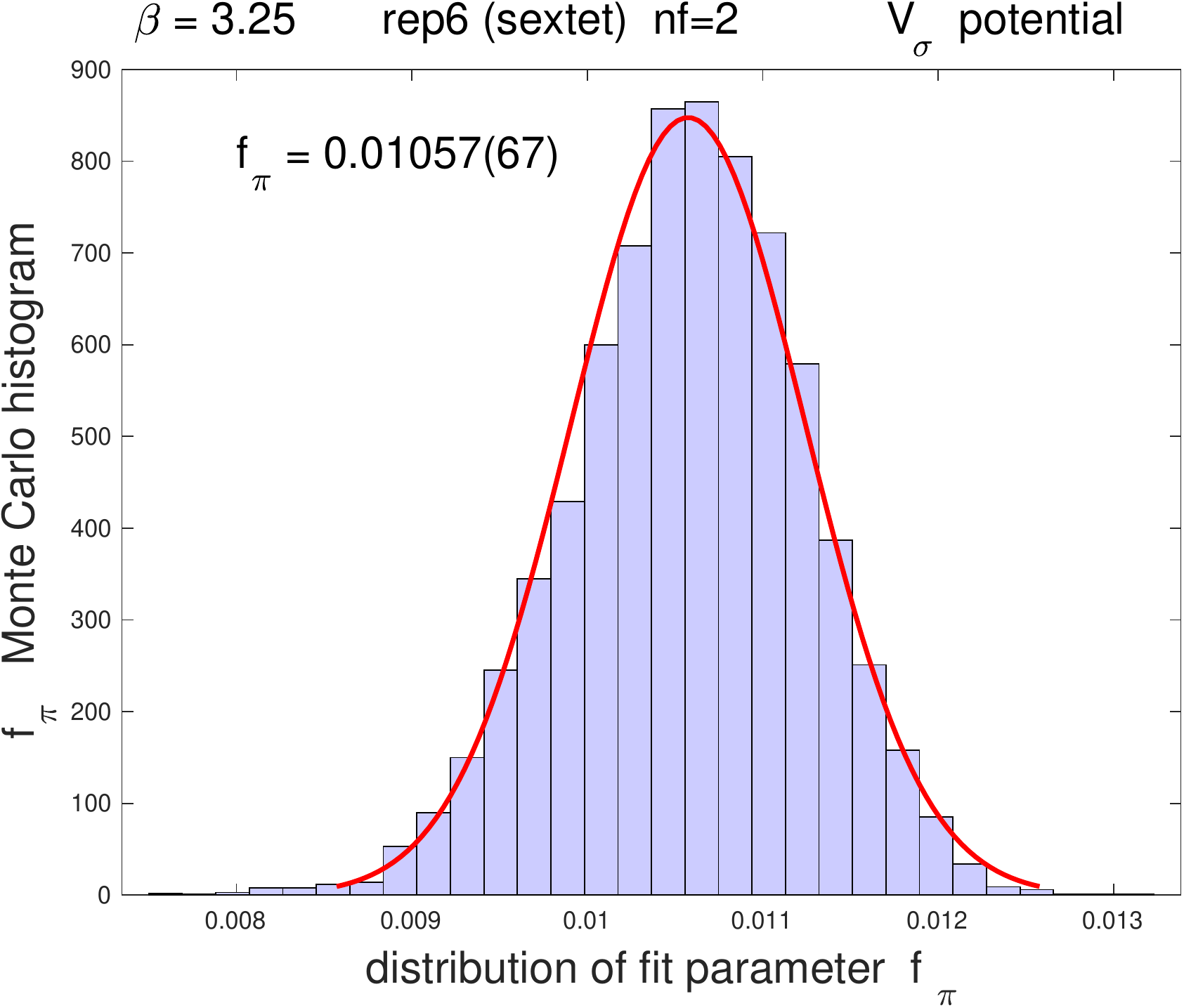}
			\end{subfigure}
			\begin{subfigure}{.49\linewidth}
				\caption*{\footnotesize posterior $\Sigma(0)=B_\pi\cdot f_\pi^2$  :}
				\includegraphics[width=\textwidth]{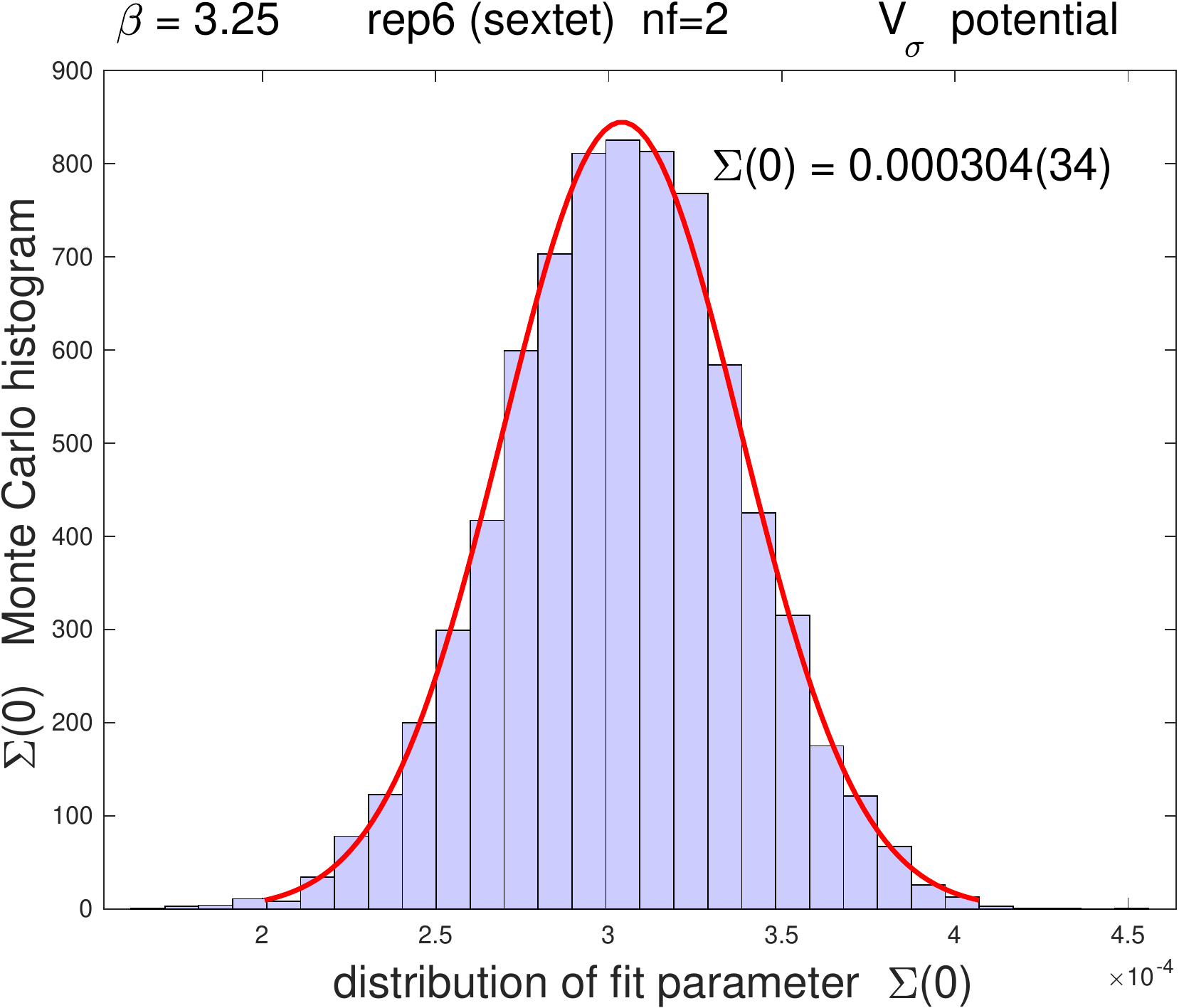}
				%			\caption{\footnotesize fit result for $B_\pi$}
			\end{subfigure}\\
			\begin{subfigure}{.49\linewidth}
				\caption*{\footnotesize posterior $\gamma$ :}
				\includegraphics[width=\textwidth]{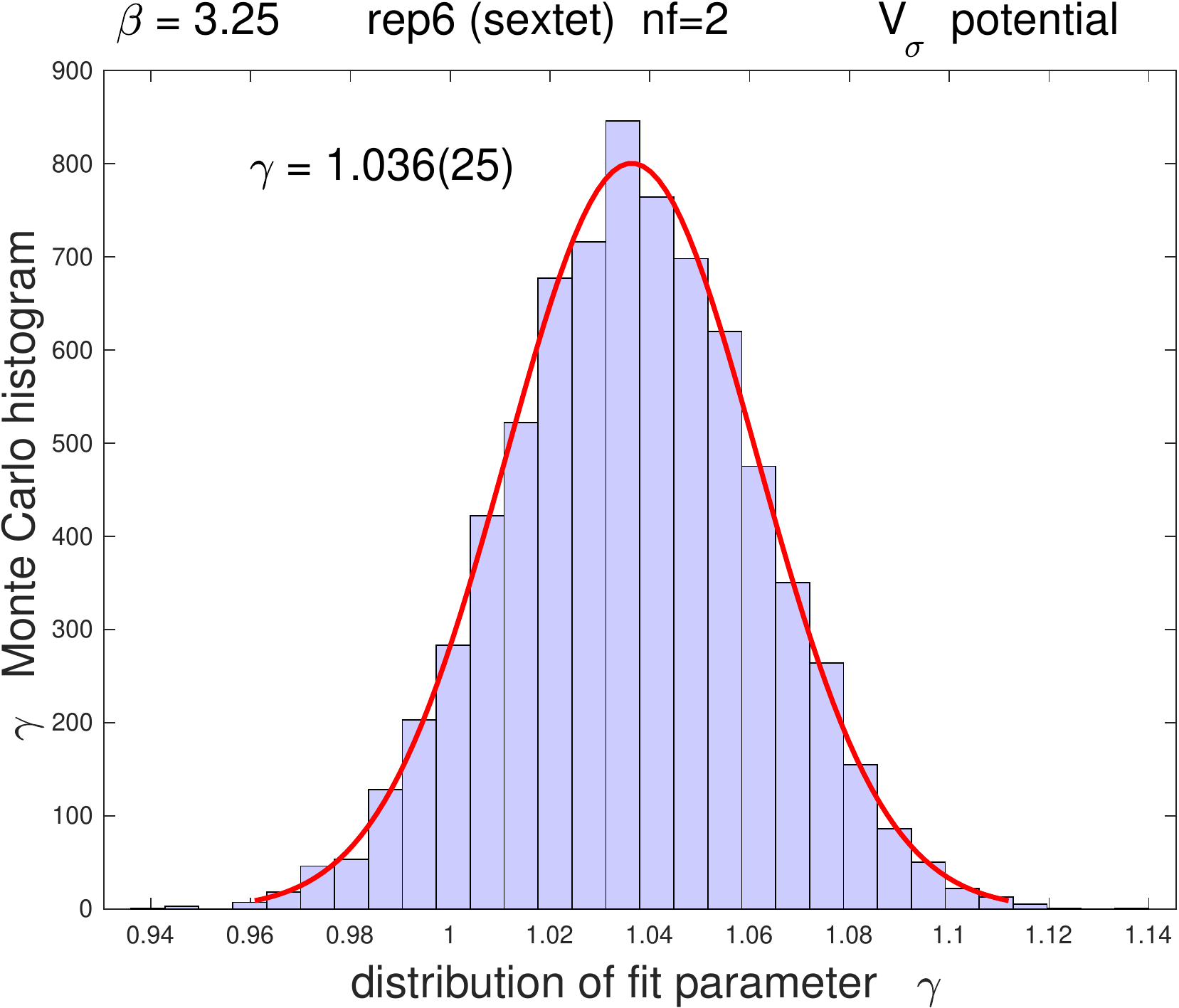}
			\end{subfigure}
			\begin{subfigure}{.49\linewidth}
				\caption*{\footnotesize posterior $(m_d/f_\pi)\cdot (f_d/f_\pi)$ :}
				\includegraphics[width=\textwidth]{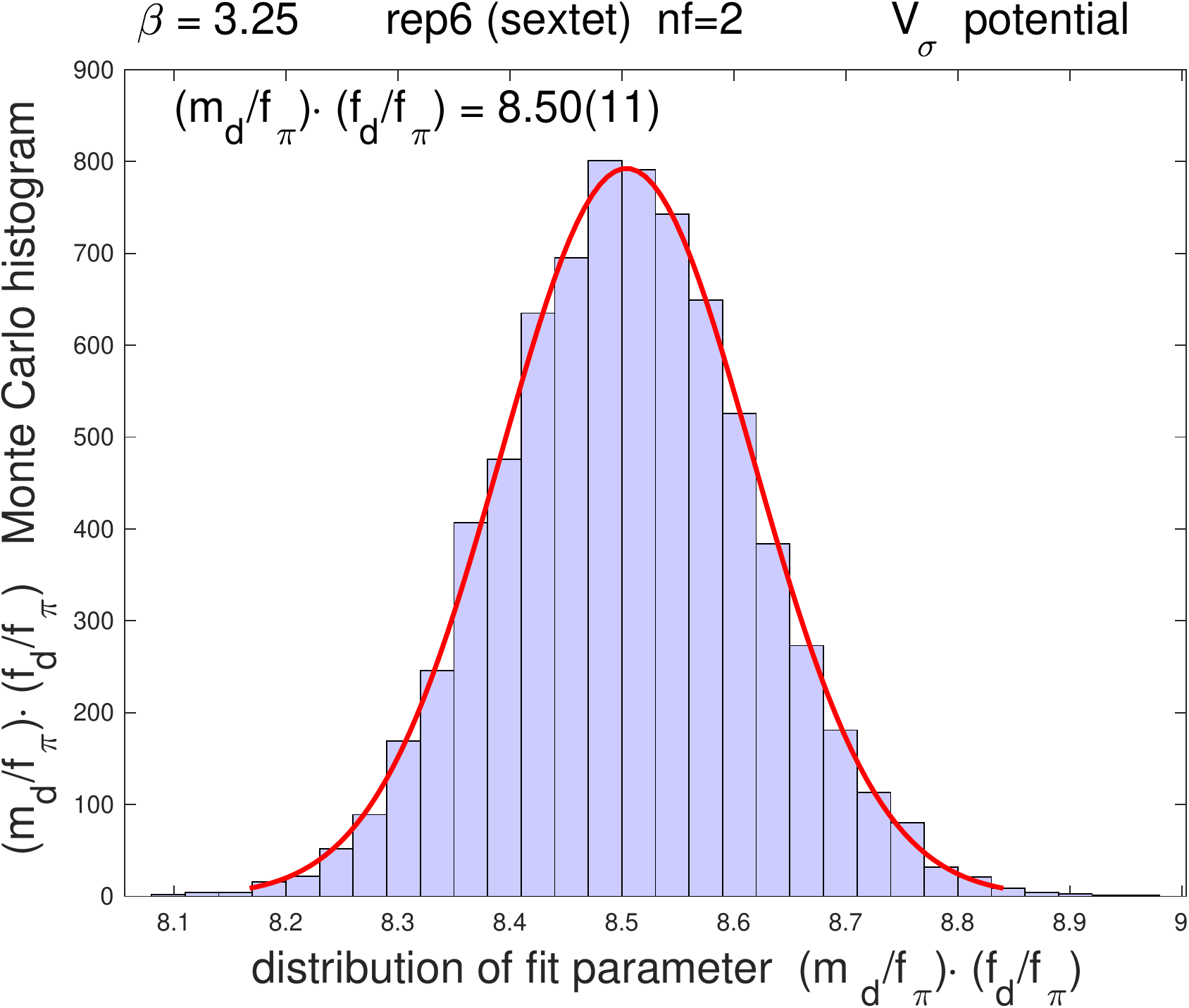}
			\end{subfigure}
		}
		\begin{subfigure}{.44\textwidth}
			\includegraphics[width=\textwidth]{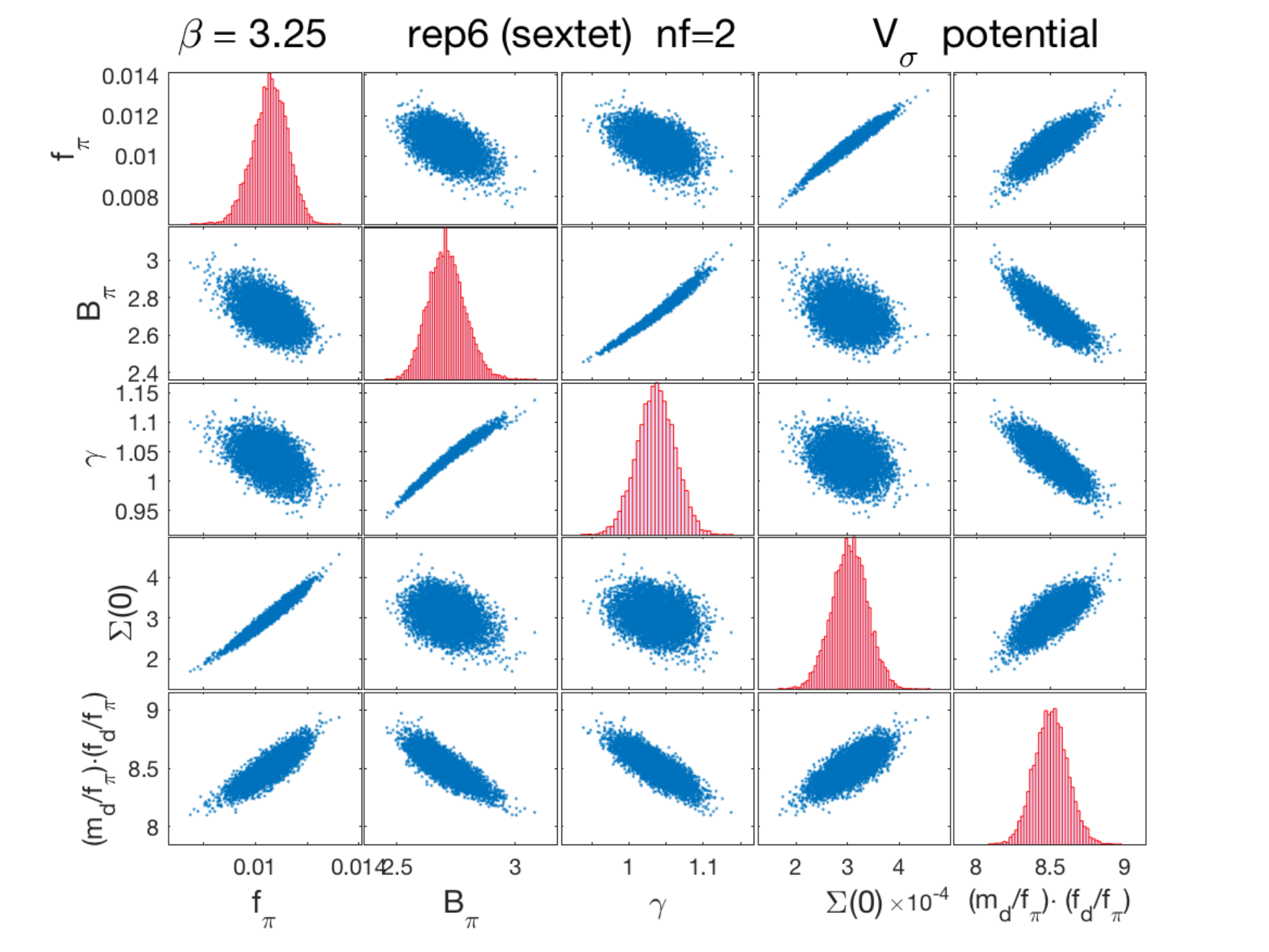}
			\caption*{\footnotesize Matrix plot of five fitted physical parameters with posterior histogram in the diagonal 
				and off-diagonal scatter plots of the correlations. Four of the histograms are also shown on the left with fitted means and
				1$\sigma$ equivalent percentile errors (the distributions are close to normal).
			}
		\end{subfigure}\hfill
		\caption{\footnotesize The four independent combinations of the five parameters without $M_d(m)$ input to  Eq.~(\ref{eq:EFT}) were determined from the posterior 
			distributions  of the  exact IML based analysis and illustrated on the four left panels where $m_d/f_\pi$ and $f_d/f_\pi$ are replaced 
			by their product.  Average values and 1$\sigma$ percentile errors are consistent with normal distributions, displayed by solid red line fits. The histogram
			with average value $B_\pi = 2.711(77)$ is not shown.
			The $\chi^2 < 1$ average of the posterior $\chi^2$ distribution, not shown, indicates a consistent fitting procedure for the $V_\sigma$ dilaton potential.
		}
		\label{fig:rep6Vsigma325}
	\end{figure}
	\vskip -0.2in

	\begin{figure}[h!]
	\centering
	\parbox{.53\textwidth}{
		\begin{subfigure}{.49\linewidth}
			\caption*{\footnotesize posterior $f_\pi$ :}
			\includegraphics[width=\textwidth]{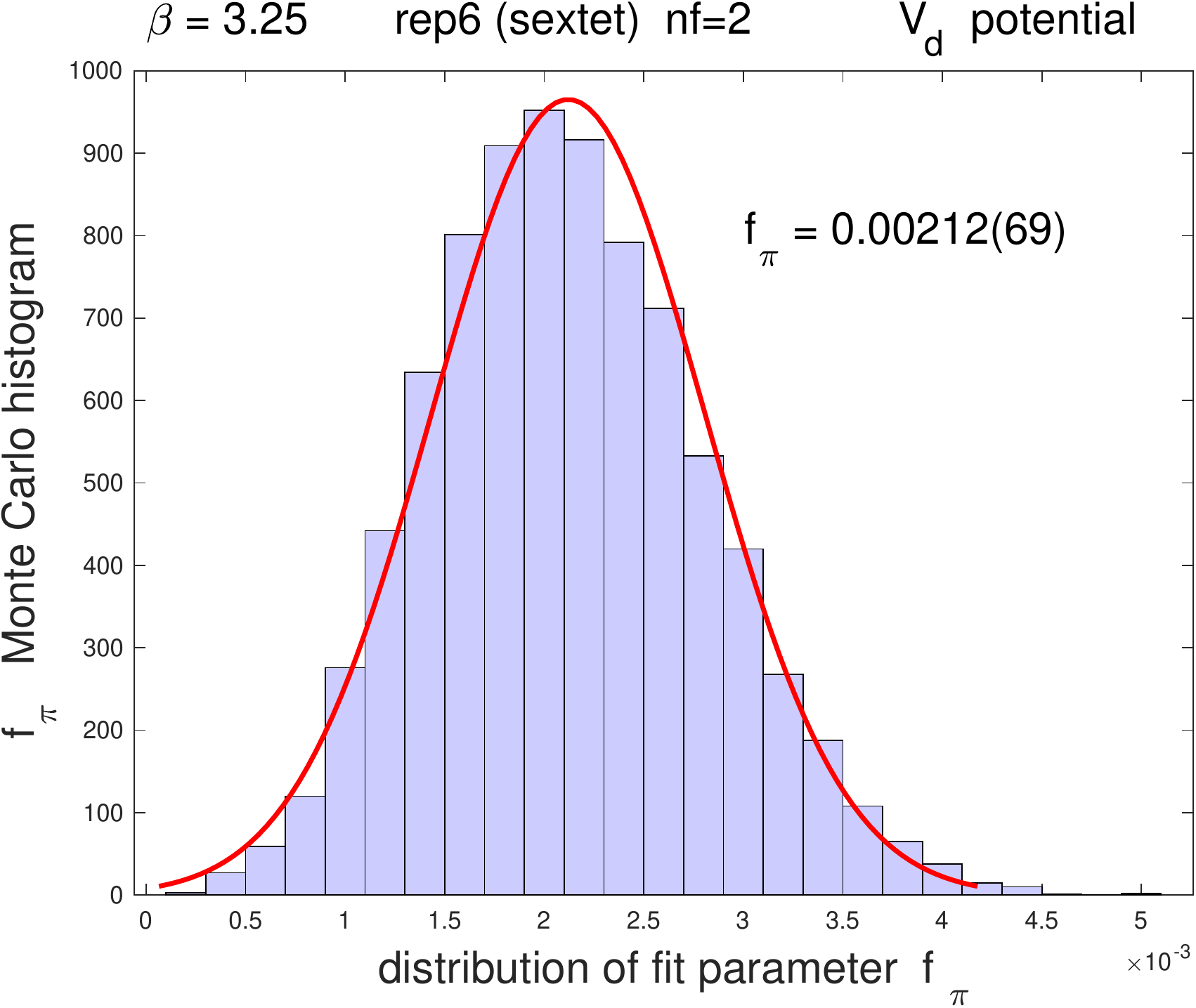}
		\end{subfigure}
		\begin{subfigure}{.49\linewidth}
			\caption*{\footnotesize posterior  $\Sigma(0)=B_\pi\cdot f_\pi^2$ :}
			\includegraphics[width=\textwidth]{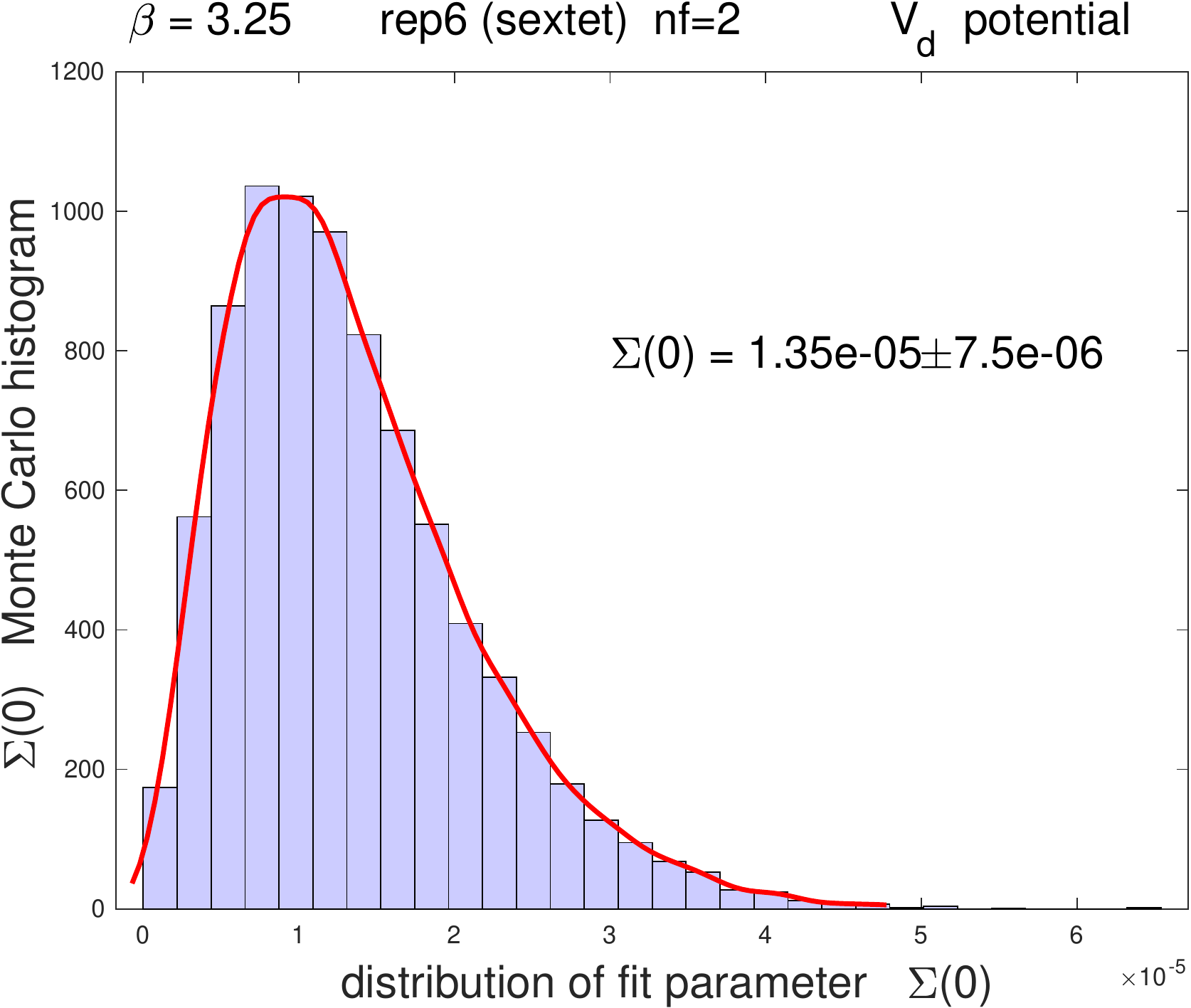}
		\end{subfigure}\\
		\begin{subfigure}{.49\linewidth}
				\caption*{\footnotesize posterior $\gamma$  :}
				\includegraphics[width=\textwidth]{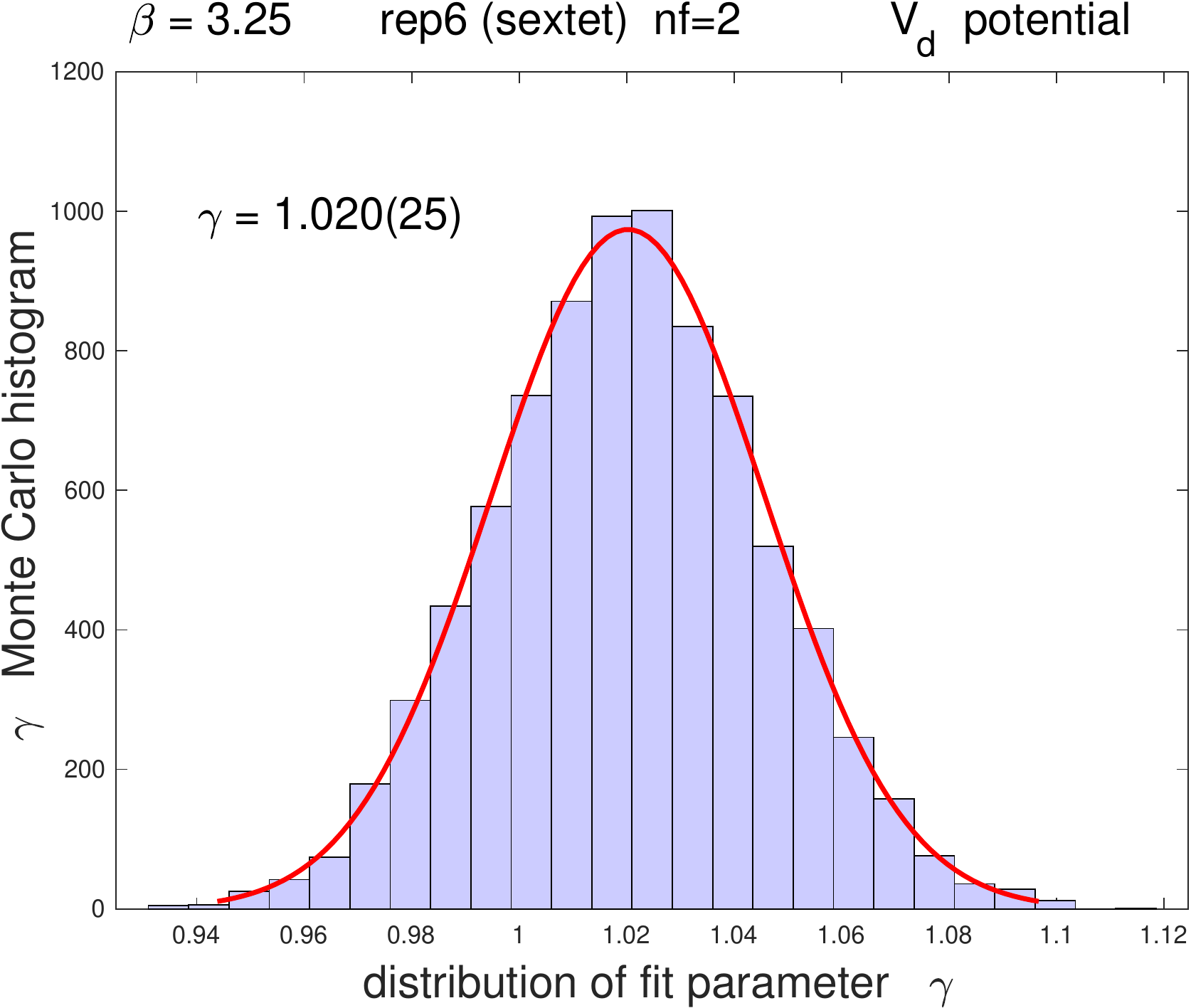}
		\end{subfigure}
		\begin{subfigure}{.49\linewidth}
	  			\caption*{\footnotesize posterior $(m_d/f_\pi)\cdot (f_d/f_\pi)$  :}
				\includegraphics[width=\textwidth]{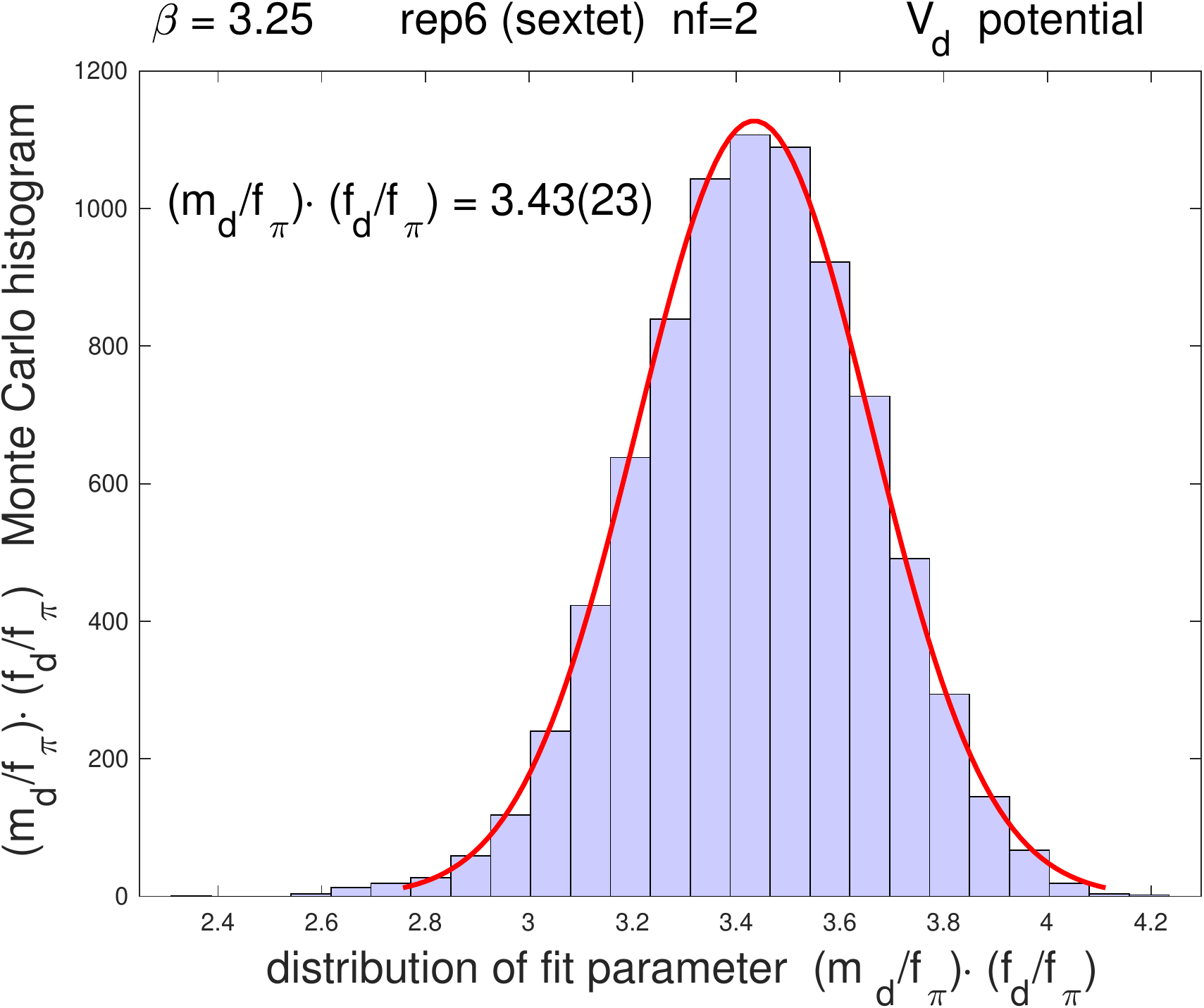}
		\end{subfigure} 
}
		\begin{subfigure}{.44\textwidth}
			\includegraphics[width=\textwidth]{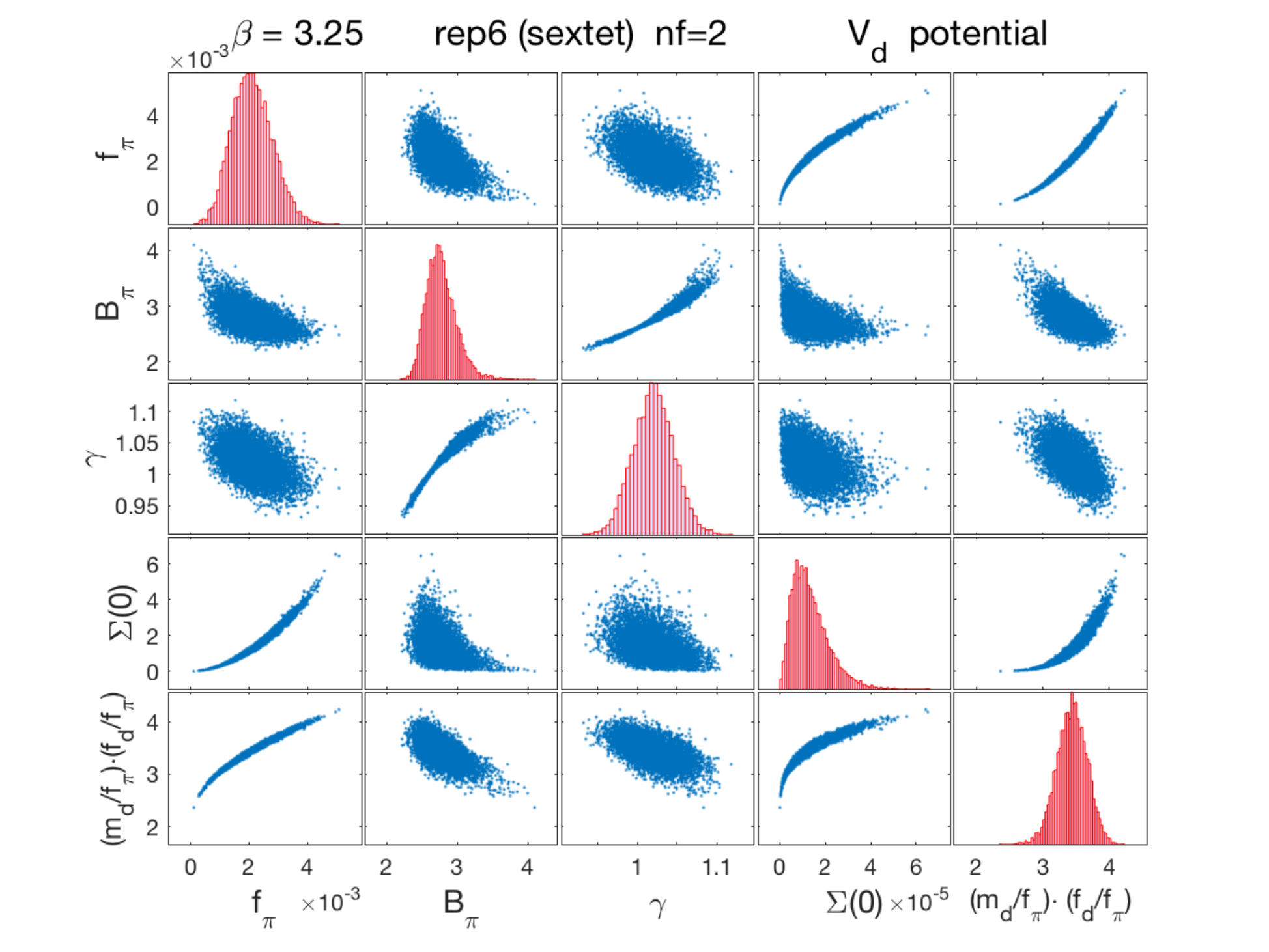}
			\caption*{\footnotesize Matrix plot of five fitted physical parameters with their posterior histograms in the diagonal 
				and off-diagonal scatter plots of their correlations. Four of the histograms are also shown on the left with fitted means and
				1$\sigma$ equivalent percentile errors (the distributions are close to normal, except $\Sigma(0)$).}
		\end{subfigure}\hfill
    	
%		\vskip -0.1in
		\caption{\footnotesize  The four independent combinations of the five parameters without $M_d(m)$ input to  Eq.~(\ref{eq:EFT}) were determined from the posterior 
			distributions  of the  exact IML based analysis and illustrated on the four left panels where $m_d/f_\pi$ and $f_d/f_\pi$ are replaced 
			by their product.  Average values and 1$\sigma$ percentile errors are consistent with normal distributions, displayed by solid red line fits. The histogram
			with average value $B_\pi = 2.77(21)$ is not shown.
			The $\chi^2 < 1$ average of the posterior $\chi^2$ distribution, not shown here, would indicate a consistent fitting procedure for the $V_d$ dilaton potential
			but the extreme low value of $f_\pi$ casts doubts on the LO $V_d$ analysis of Eq.~(\ref{eq:EFT}. }
		\label{fig:rep6Vd325}
	\end{figure}
	\newpage
	
	%%%%%%%%%%%%%%%%%%%%%%%%%%%%%%%%%%%%%%%%%%%%%
	
	\noindent{\bf 4.4 ~~ Results in the p-regime on $\mathbf{f_\pi}$ and the chiral condensate from the two dilaton potentials}
	
	Tests based on LO p-regime analysis of the $V_\sigma$ dilaton potential show  reasonable consistency as the lattice spacing is varied with the  lattice scale at gauge coupling $\beta = 3.20$ set by $t_0/a^2 = 6.20(14)$  and  by $t_0/a^2 = 10.48(23)$ at $\beta=3.25$ (the physical scale $t_0$ was defined in our earlier work from the gauge field gradient flow at  gauge coupling $g^2_R = 6.7$ in the chiral limit~\cite{Fodor:2017wsn}). Even at the smaller lattice spacing, $f_\pi\cdot L$ is not far from being $\mathcal{O}(1)$ and realistic to increase in future work, as required in the p-regime. The observed drop $f^{3.20}_\pi/f^{3.25}_\pi = 1.34$, as read from Fig.~\ref{fig:rep6Vsigma325} at the $\beta=3.25$  lattice scale, is close to the drop $a_{3.20}/a_{3.25}=1.30$ from $t_0$ scale setting. Similarly, in the GMOR limit the drop $\Sigma(0)_{3.20}/\Sigma(0)_{3.25} = 2.36$ from Fig.~\ref{fig:rep6Vsigma325} is close to the drop $(a_{3.20}/a_{3.25})^3=2.20$  from $t_0$ scale setting (the a-dependent renormalization constant $Z_p(a,\mu)$ is needed for more refined scaling tests of the physical $\Sigma$-condensate).
	
     In contrast with the $V_\sigma$ p-regime fit of $f^\sigma_\pi=0.01413(58)$ for the fundamental parameter $f_\pi$ at $\beta=3.20$, we find  the surprisingly low  $f^d_\pi=0.00302(69)$ fit from the $V_d$ potential at the same gauge coupling, predicting the GMOR condensate $\Sigma_{d}(0)$ to be very low compared to $\Sigma_{\sigma}(0)$.
	 A similar pattern emerged for the two dilaton potentials from the p-regime fits of $f_\pi$  in the $n_f=8$ model~\cite{Fodor:2019vmw} and confirmed again from the exact IML analysis at the conference (Wong's talk) without including the fits  again in this report. For reducing uncertainties in fitting fundamental parameters, defined in the chiral limit but fitted high above it in the p-regime, RMT analysis in the $\epsilon$-regime is the natural choice for direct determination of $f_\pi$ and $\Sigma(0)$. 
	 We will show in Section 5 that within the required condition of $f^\sigma_\pi\cdot L \gtrsim 1$  in the $\epsilon$-regime, $\Sigma(0)$ from the RMT analysis is consistent with the p-regime prediction $\Sigma_{\sigma}(0)$  from the $V_\sigma$ potential. Self-consistent analysis of the $V_\sigma$ potential would render the  $V_d$ potential inconsistent, even without RMT tests on supersized lattices under the unrealistic $f^d_\pi\cdot L \gtrsim 1$ condition.  Further tests would be important to confirm the self-consistency of predictions from the $V_\sigma$ potential, including the direct test of the p-regime fit $f^\sigma_\pi=0.01413(58)$  in the chiral limit from RMT analysis of the lowest Dirac eigenvalues when imaginary chemical potential is introduced in the analysis. This investigation is ongoing with our extended code suite.  
	 Based on theoretical arguments, the $V_d$ potentials represents an attractive dilaton EFT. Difficulties with its first tests from RMT studies in the $\epsilon$-regime will stimulate further systematics and statistics, like the challenging taste breaking effects in the dynamical RMT spectrum, as reported in Section 5.

%%%%%%%%%%%%%%%%%%%%%%%%%%%%%%%%%%%%%%%%%%%%%
	
	%
	\section{Pilot study of the dilaton EFT in the $\mathbf{\epsilon}$-regime from RMT Dirac spectra}\label{section:rmt}
	\vskip -0.1in
	\noindent{\bf 5.1 ~~ The  leading order ${ {\cal Z}^{LO}_\epsilon(m) }$ partition function: }In the $\epsilon$-regime, close to the chiral limit where the pion correlation length far exceeds the 
	linear size of the finite volume, the EFT Lagrangian density of Eq.~(\ref{eq:EFT}) simplifies to
	\begin{equation}
	{\cal L_\epsilon} = \frac{1}{2}\partial_{\mu} \chi \partial_{\mu} \chi\,-\,V_{\sigma,d} (\chi) +
	\frac{m^2_\pi f^2_\pi}{4}\big(\frac{\chi}{f_d}\big)^y ~ {\rm tr}\big[U_0 + U_0^\dagger\big], \label{eq:EFT1}
	\end{equation}
	as we argued in~\cite{Fodor:2019vmw}, applicable  to both choices $V_{\sigma/d}$ of the dilaton potential. In Eq.~(\ref{eq:EFT1}) the coupling of the dilaton to the  zero momentum mode $U_0$  of $SU(2)$ pion dynamics is represented by the $\chi(x)$  field.\footnote {We changed the notation of~\cite{Fodor:2019vmw} from $\Sigma_0$ to $U_0$ to minimize confusion with  $\Sigma(0)$ designating the chiral condensate in this report.} The fluctuations of the non-zero momentum modes of the dilaton field can be treated by systematic expansion as required by RMT applications~\cite{Fodor:2019vmw}.   Close to the chiral limit and based on the results of the p-regime analysis, we generated  lattice ensembles where the fluctuations of the dilaton field are treated as quenched in the p-regime, like for any other massive state.  Only the zero momentum mode of the dilaton participates  with a  shift  to the new minimum  $f^{\epsilon}_d(m)$  from $f_d $ in the dilaton potential, as driven by small fermion mass deformations in the LO EFT partition function ${\cal Z}^{LO}_\epsilon(m)$,
	\begin{equation}
		{\cal Z}^{LO}_\epsilon(m) ={\rm  exp}\left[-{\rm V}\cdot V_{\sigma/d}(f^{\epsilon}_d(m))\right]\cdot\int_{SU(2)} DU_0 ~{\rm exp}\left[
		\frac{m^2_\pi f^2_\pi}{4}\left(\frac{f^{\epsilon}_d(m)}{f_d}\right)^y  V~ {\rm Tr}\left(U_0 + U_0^\dagger\right)\right] , \label{eq:epsilon}
	\end{equation}
	with lattice volume $V$ for either choice of the $V_{\sigma/d}(f^{\epsilon}_d(m))$ dilaton potential at the minimum. The notation was introduced in the previous section with one important difference. While in the p-regime  the shift in the dilaton potential from $f_d$ to $F_d(m)$ in the presence of fermion mass deformations was significant, the shift from $f_d$ to $f^{\epsilon}_d(m)$ in the $\epsilon$-regime is practically negligible when based on Eq.~(\ref{eq:epsilon}) for the tiny fermion mass deformation $m=0.000010$ of our lattice ensembles in the $\epsilon$-regime. Before turning to the RMT analysis of these ensembles, designed for the $\epsilon$-regime, we first summarize RMT results from our p-regime ensembles using partially quenched RMT analysis including taste breaking effects in staggered lattice fermion implementation.
	
	%%%%%%%%%%%%%%%%%%%%%%%%%%%%%%%%%%%%%%%%%%%%%
	\vskip 0.1in
    {\bf\noindent 5.2 ~~Partially quenched RMT analysis with pions and the dilaton in the p-regime}  
    
    The application of dilaton EFT to the RMT analysis of the lowest eigenvalues in the Dirac spectrum  requires two distinct sets of lattice ensembles. In the first set, also used in the p-regime analysis of the previous section, the pions and the dilaton are in the p-regime but the Dirac eigenvalues of the RMT predictions are in the $\epsilon$-regime.  This requires the application of partially quenched mixed action analysis, followed in~\cite{Bernardoni:2010nf}  and extended here to RMT analysis in the dilaton EFT of Eq.~(\ref{eq:EFT}). 
    The standard partially quenched chiral random matrix theory can be written as
    \be
    \label{RMTZ}
    \cZ^{RMT}_{N_f,N_b} =
    \int dW p(W) \frac{\prod_{f=1}^{N_f} \det(\cD_0+m_f)}{\prod_{b=1}^{N_b}
    	\det(\cD_0+m_b)} , \hskip 0.3in     	
    \cD_0 = \left( \begin{array}{cc}
    	0 & i W \\
    	i W^\dagger & 0
    \end{array} \right) ~,
    \ee
    where $W$ is a complex $N \times N$ matrix with Gaussian weight distribution p(W) (only the zero topological charge sector will be tested here from our lattice ensembles). Following the influential work of~\cite{Osborn:2010eq,Osborn:2012bc}, we added four taste breaking  RMT matrices $\cT_i~$ to the Dirac matrix, $\cD_0 + \sum_i C_i\cdot\cT_i$, with  $C_i, ~i=1,3,4,6$ coefficients  set equal from the taste breaking pattern of the sextet model at small mass deformations~\cite{Fodor:2016pls} for this pilot study.\footnote{The RMT notation for the  taste breaking matrices $\cT_i~$ and the $C_i$ coefficients is explained in~\cite{Osborn:2010eq}.} 
	\begin{figure}[h!]
		\begin{center}
			\begin{tabular}{cccc}	
				\includegraphics[height=2.9cm]{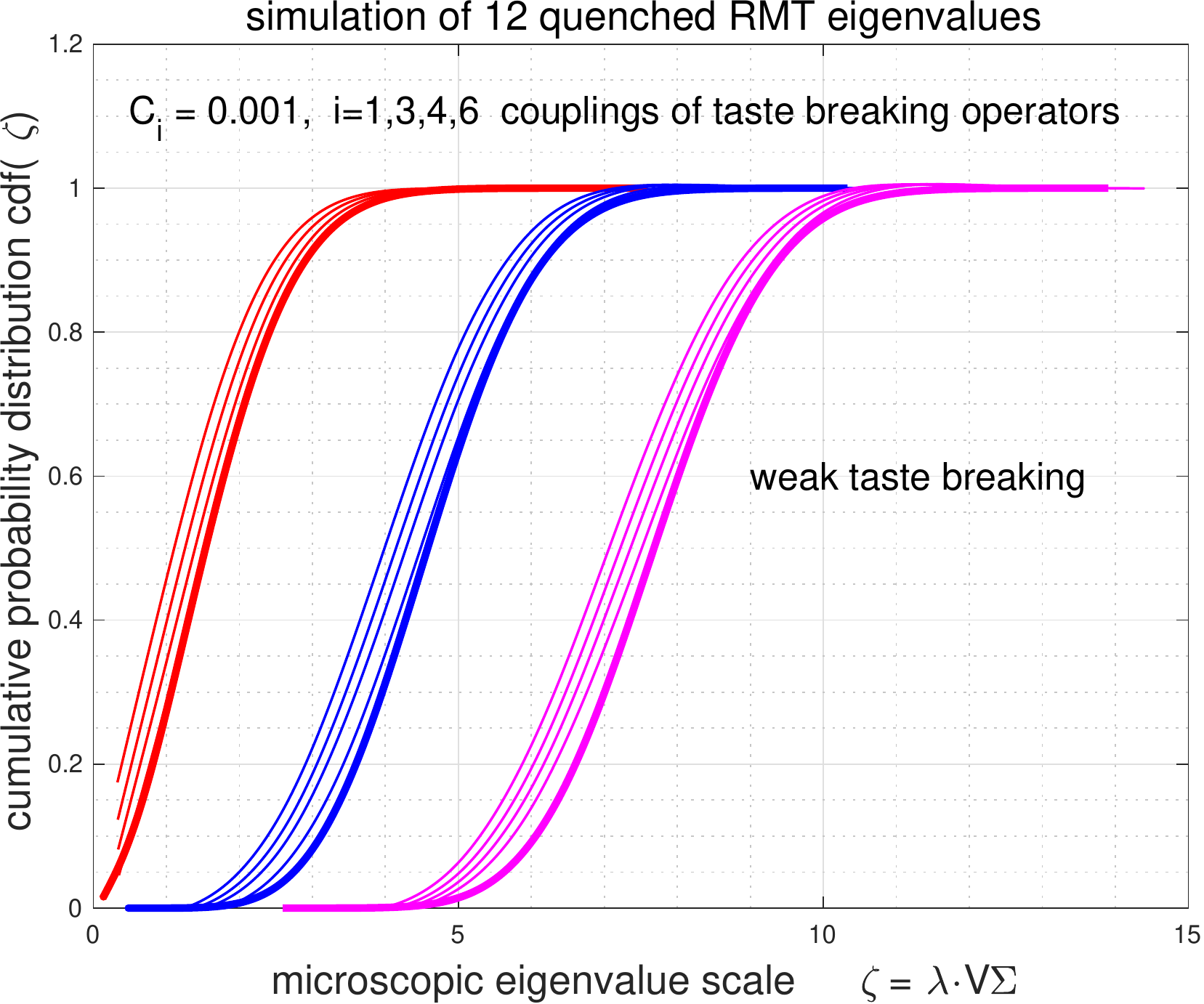}&
					\includegraphics[height=2.9cm]{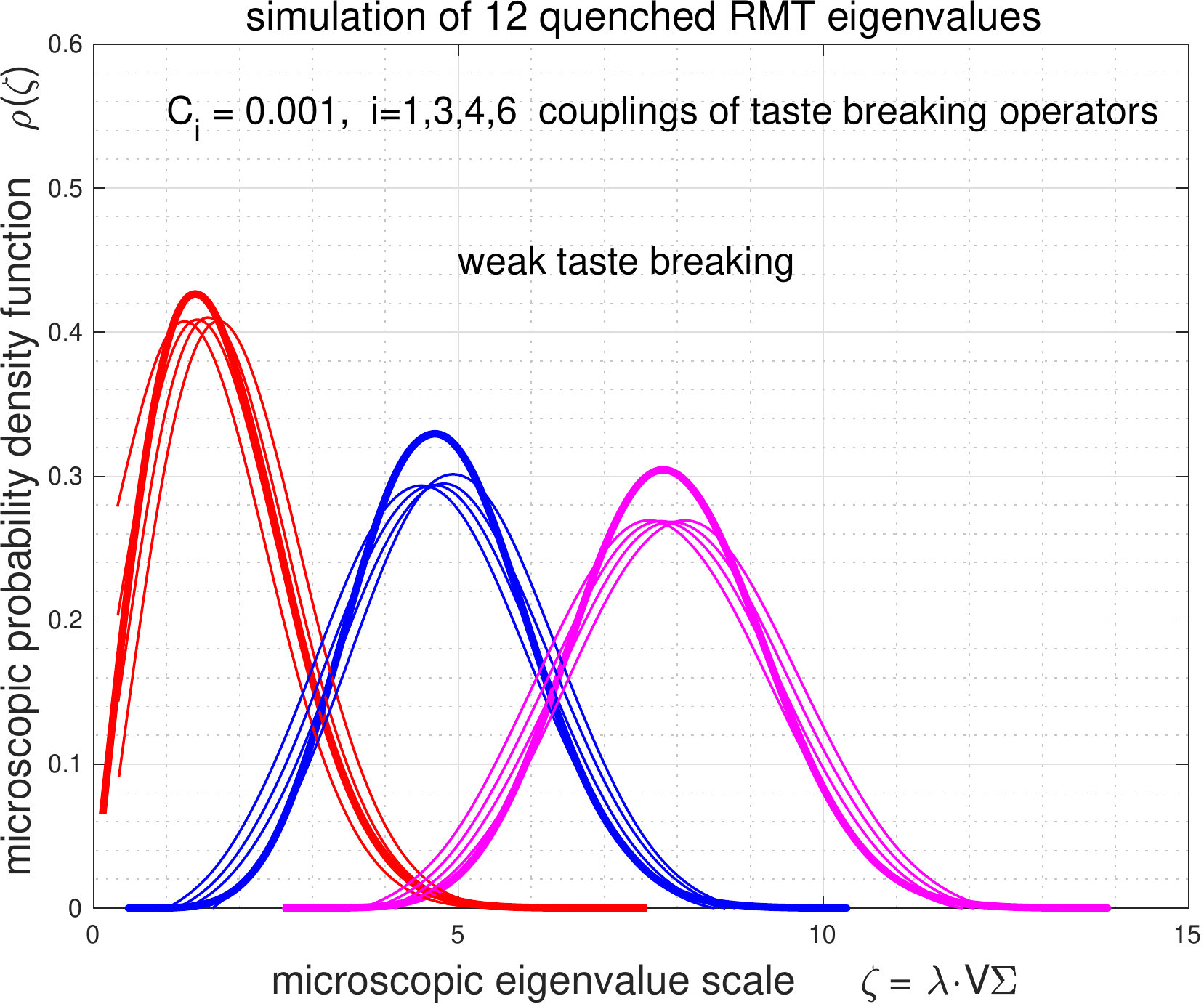}&
				\includegraphics[height=2.9cm]{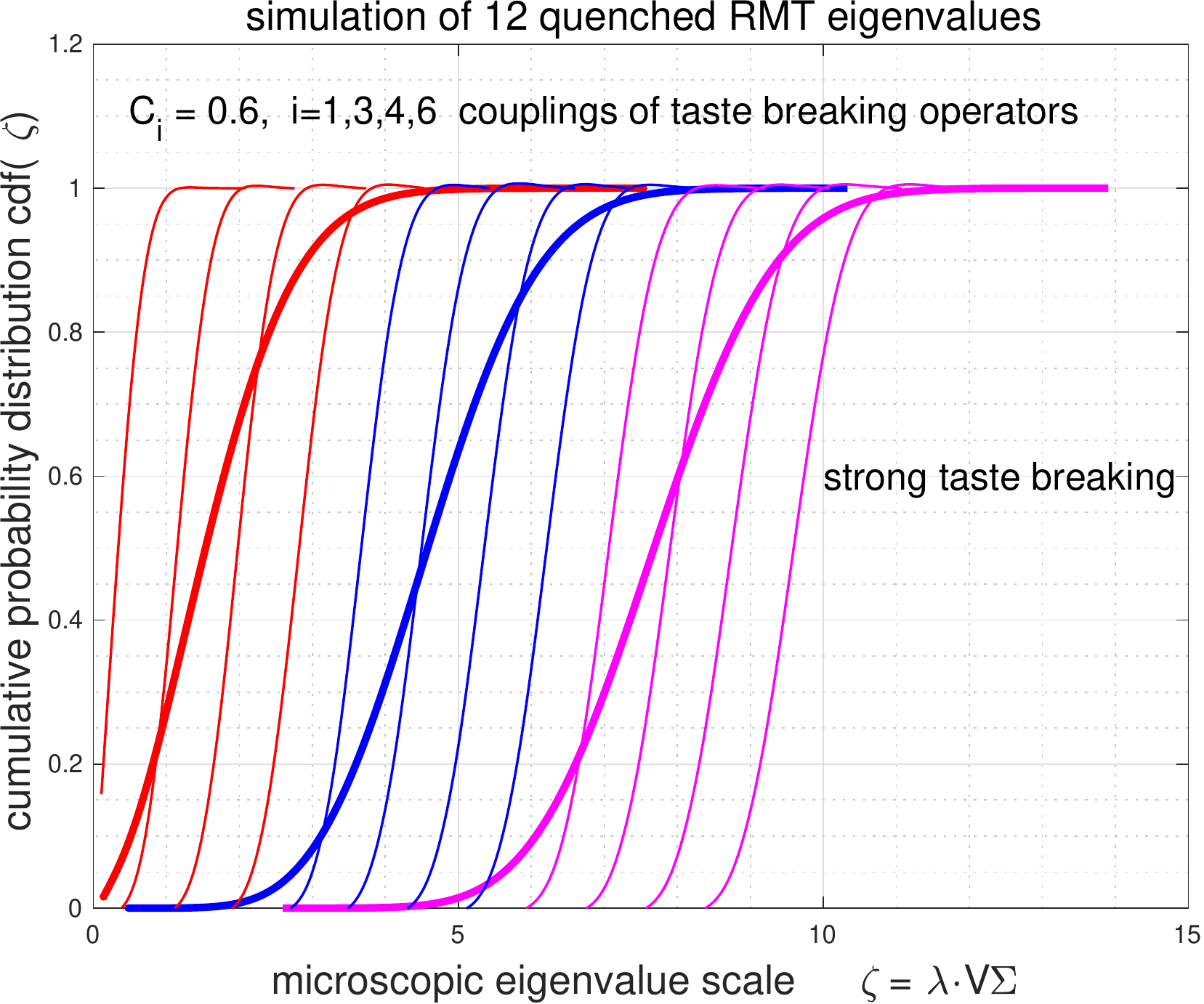}&
				\includegraphics[height=2.9cm]{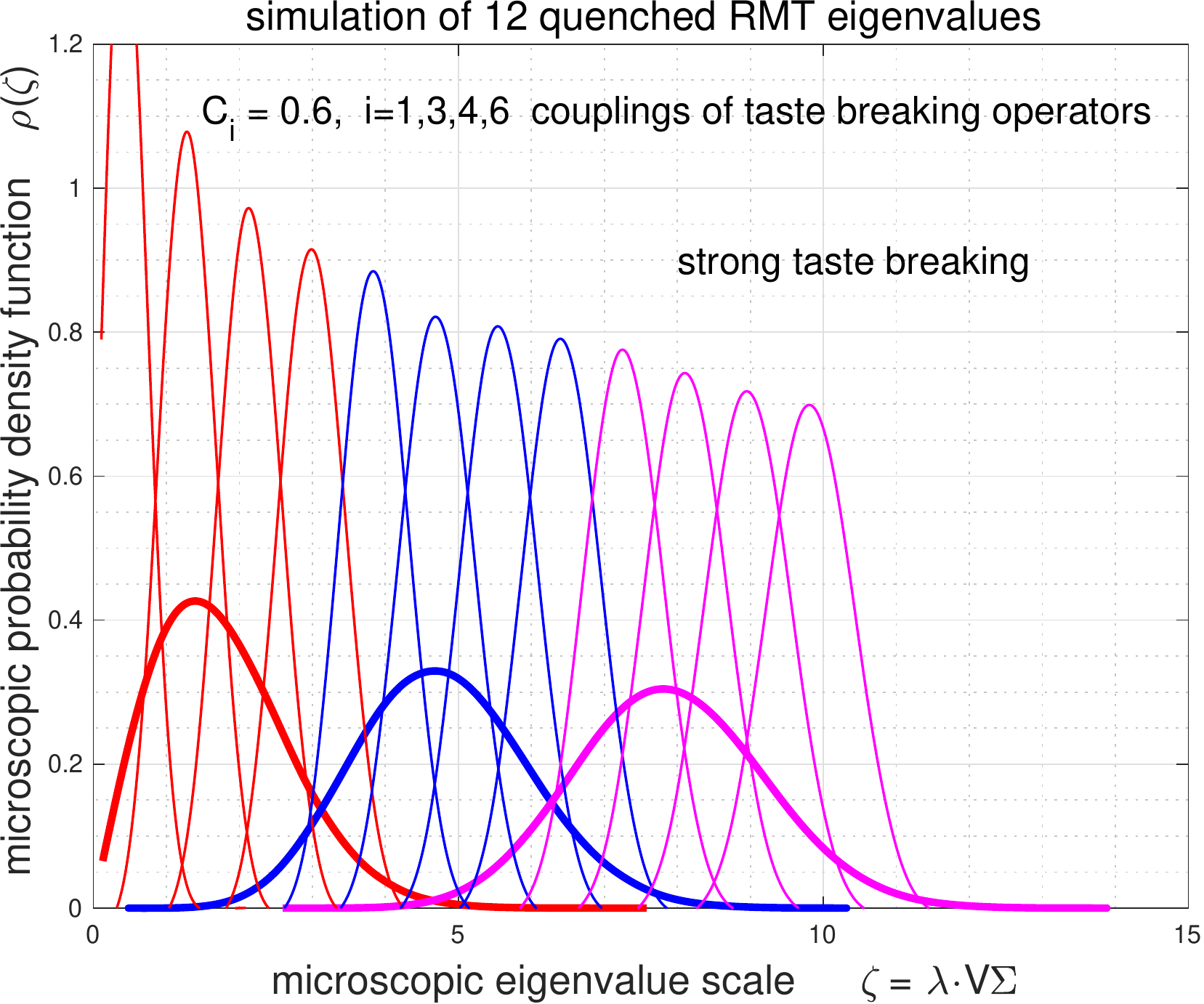}
		    \end{tabular}
		\end{center}	
	\vskip -0.25in
		\caption{\footnotesize  The three lowest quenched quartets are followed from RMT Monte Carlo simulations with crossover from  weak taste  breaking toward the U(1) symmetry breaking pattern as the $C_i$ coefficients are increased~\cite{Osborn:2010eq,Osborn:2012bc}. The thick colored lines show the degenerate eigenvalues of quartets without taste breaking and color coding tracks the split as the taste breaking $\cT_i~$ operators are turned on with $C_i$ coefficients in quenched approximation.
} 
       \vskip -0.2in
		\label{fig:rmt1}
	\end{figure}	
 To get the rather complicated RMT predictions for the lowest Dirac eigenvalues, we developed Monte Carlo simulation code for the RMT theory where the taste breaking effects, quenched and unquenched, can be numerically calculated as the $C_i, ~i=1,3,4,6$ coefficients are varied, as shown in Fig.~\ref{fig:rmt1} for the quenched theory.
 	
\begin{figure}[h!]
		\begin{center}
			\begin{tabular}{ccc}
				\includegraphics[height=4cm]{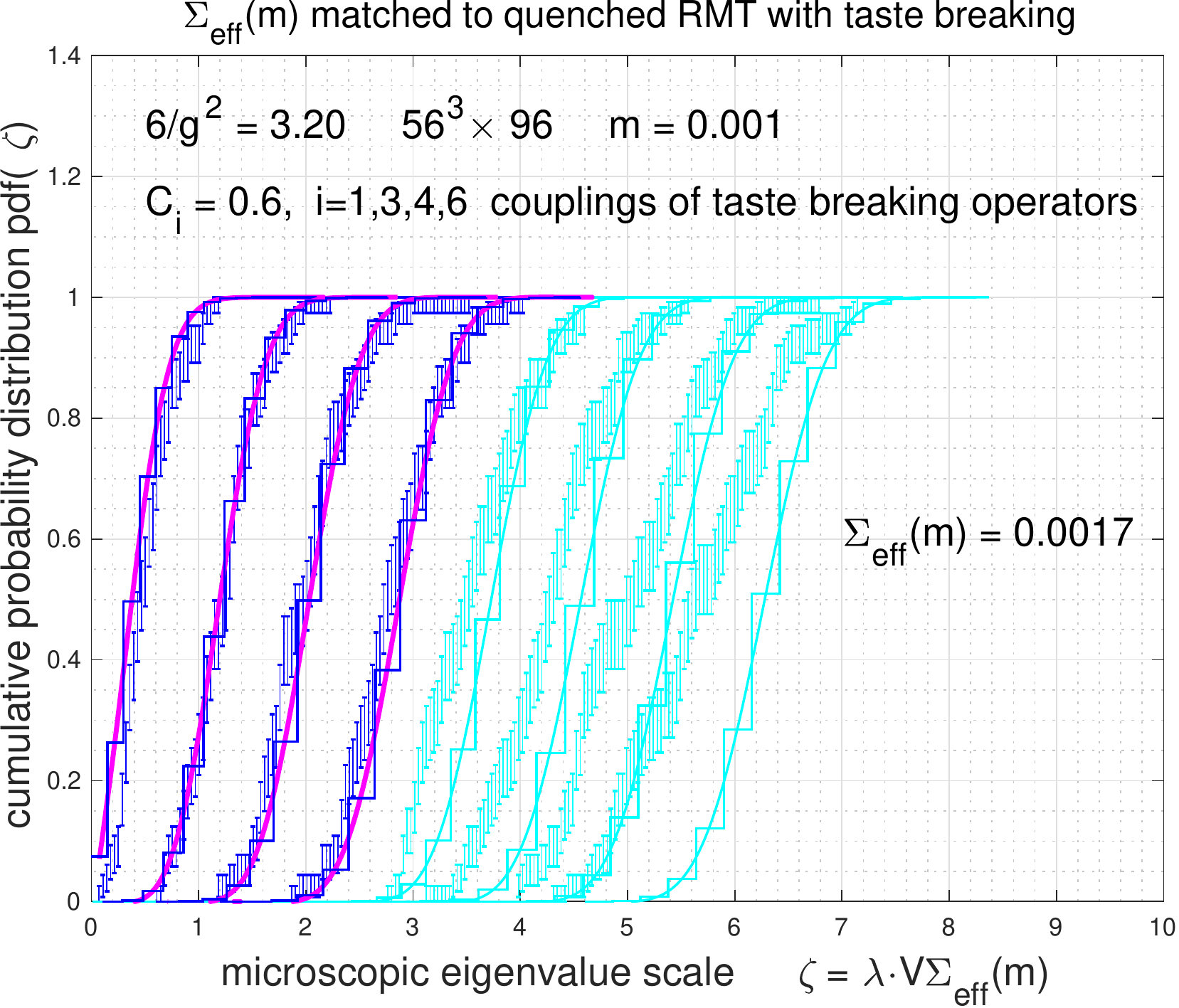}&
				\includegraphics[height=4cm]{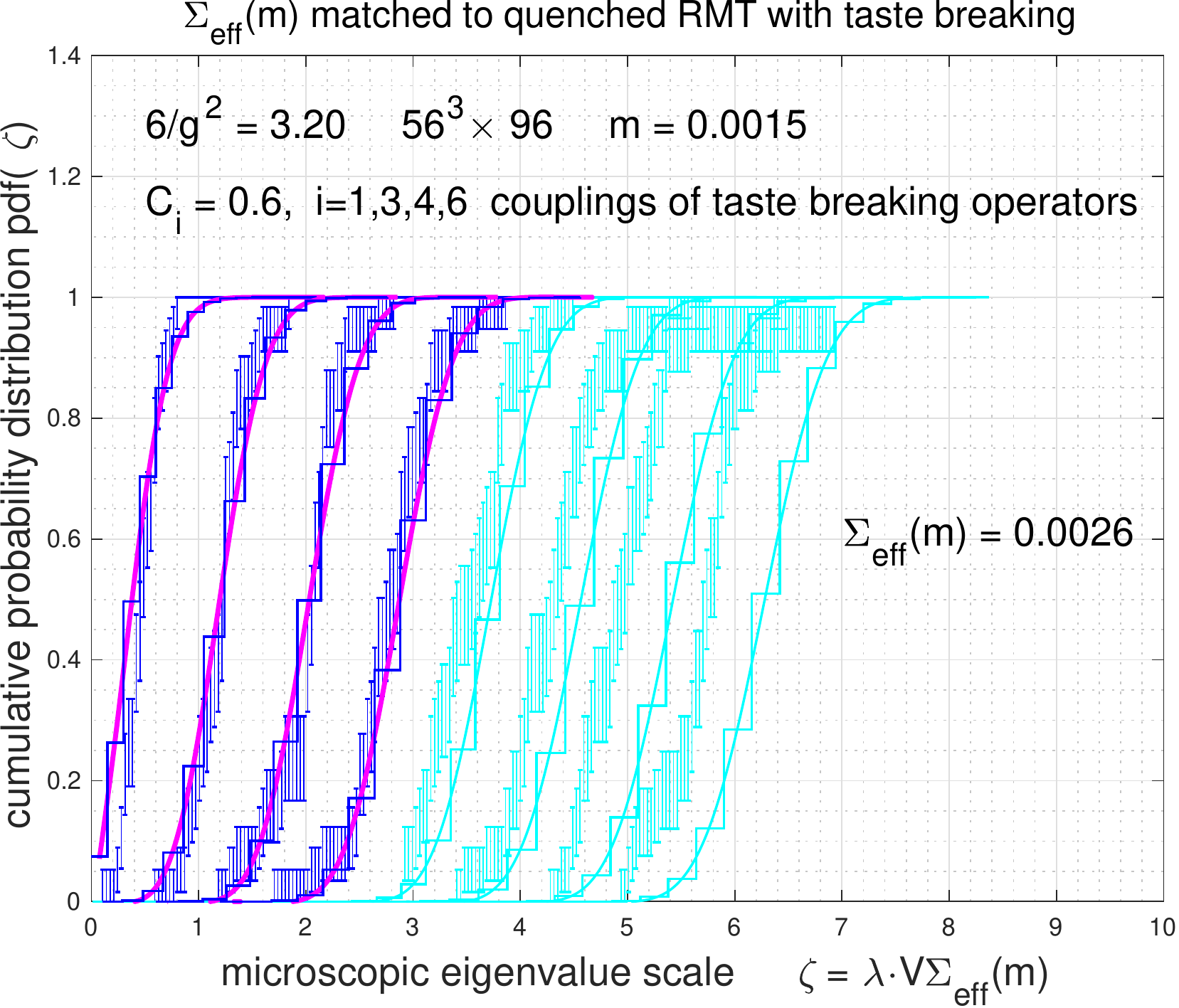}&
				\includegraphics[height=4cm]{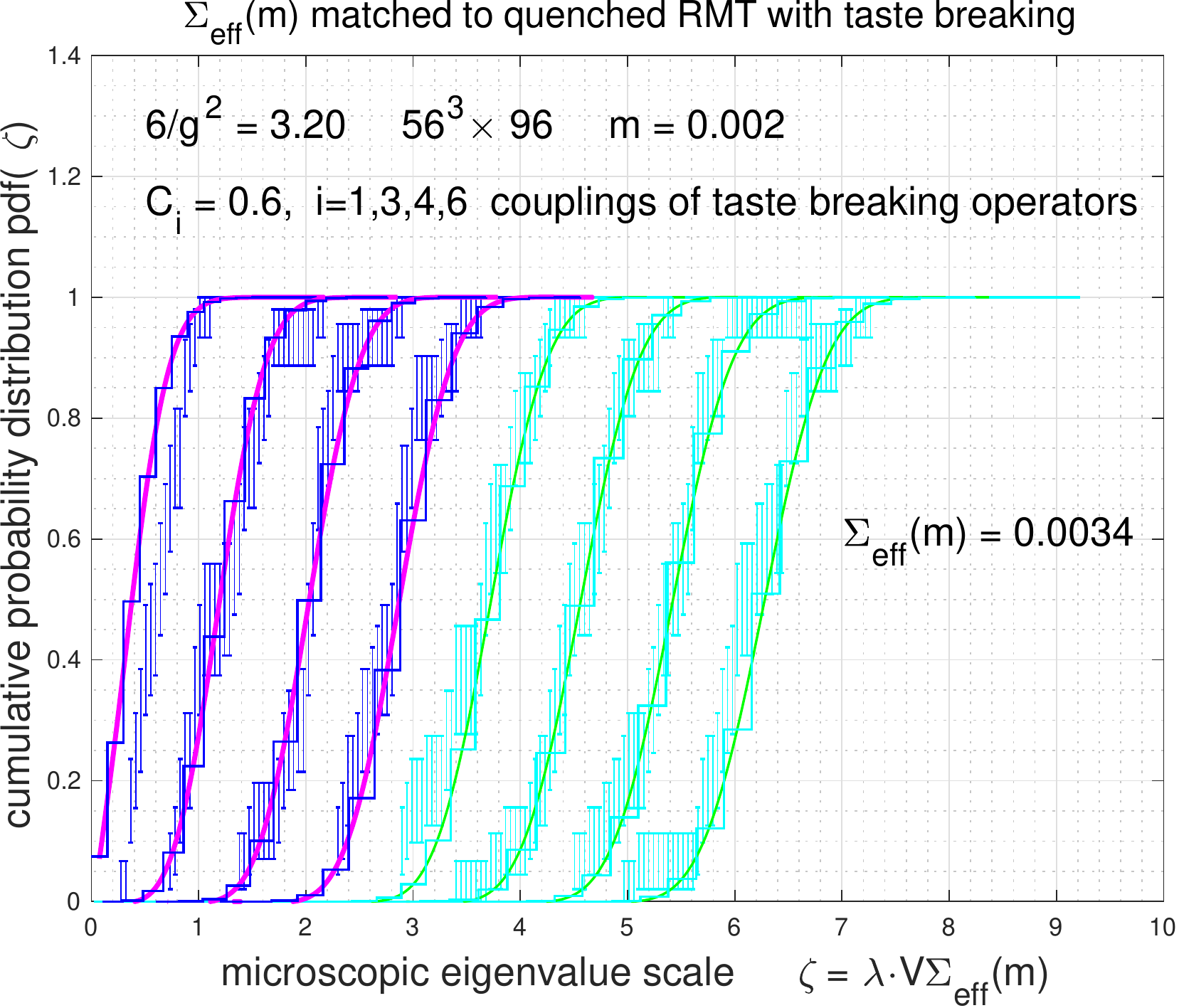}\\
				\includegraphics[height=4cm]{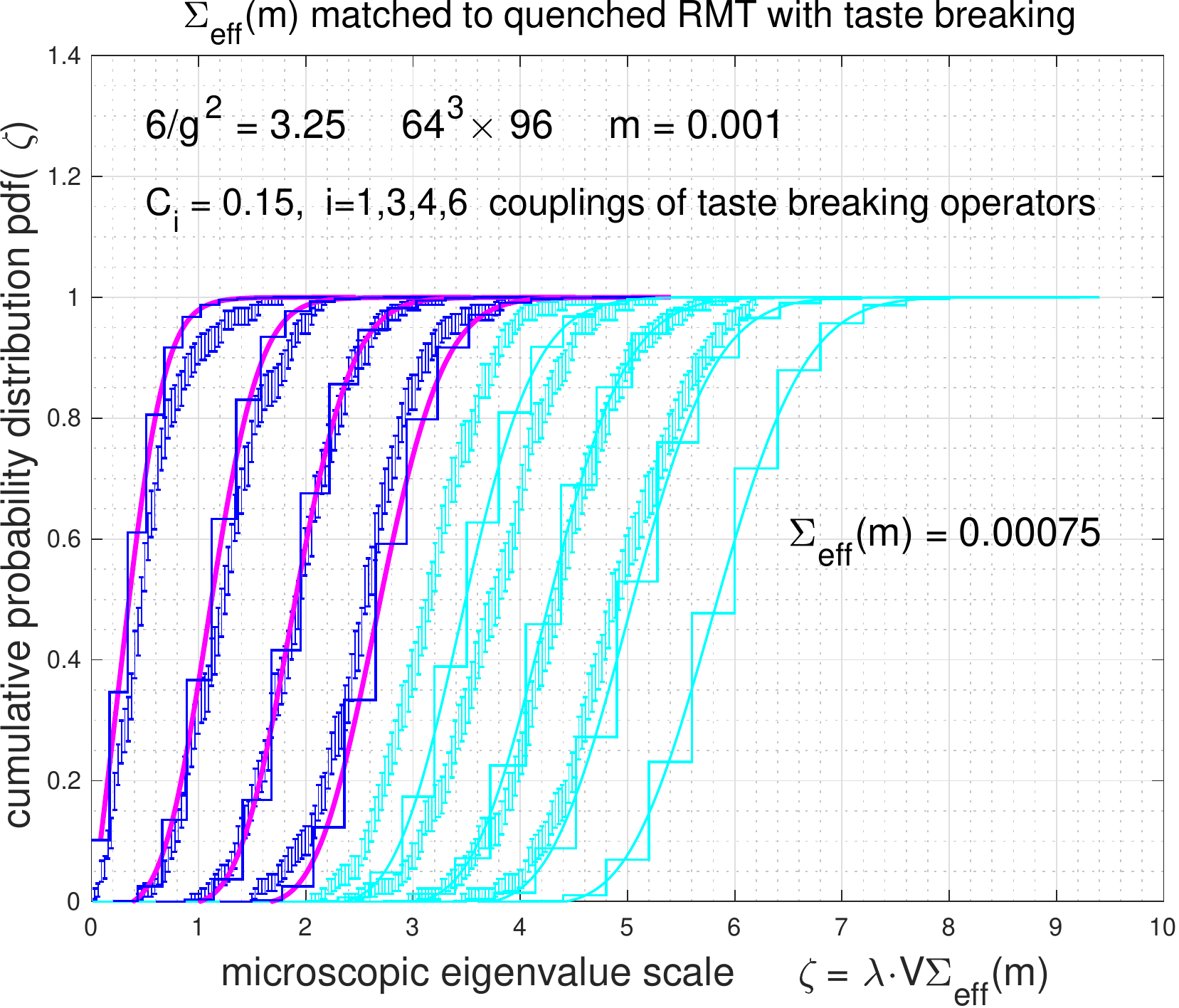}&
				\includegraphics[height=4cm]{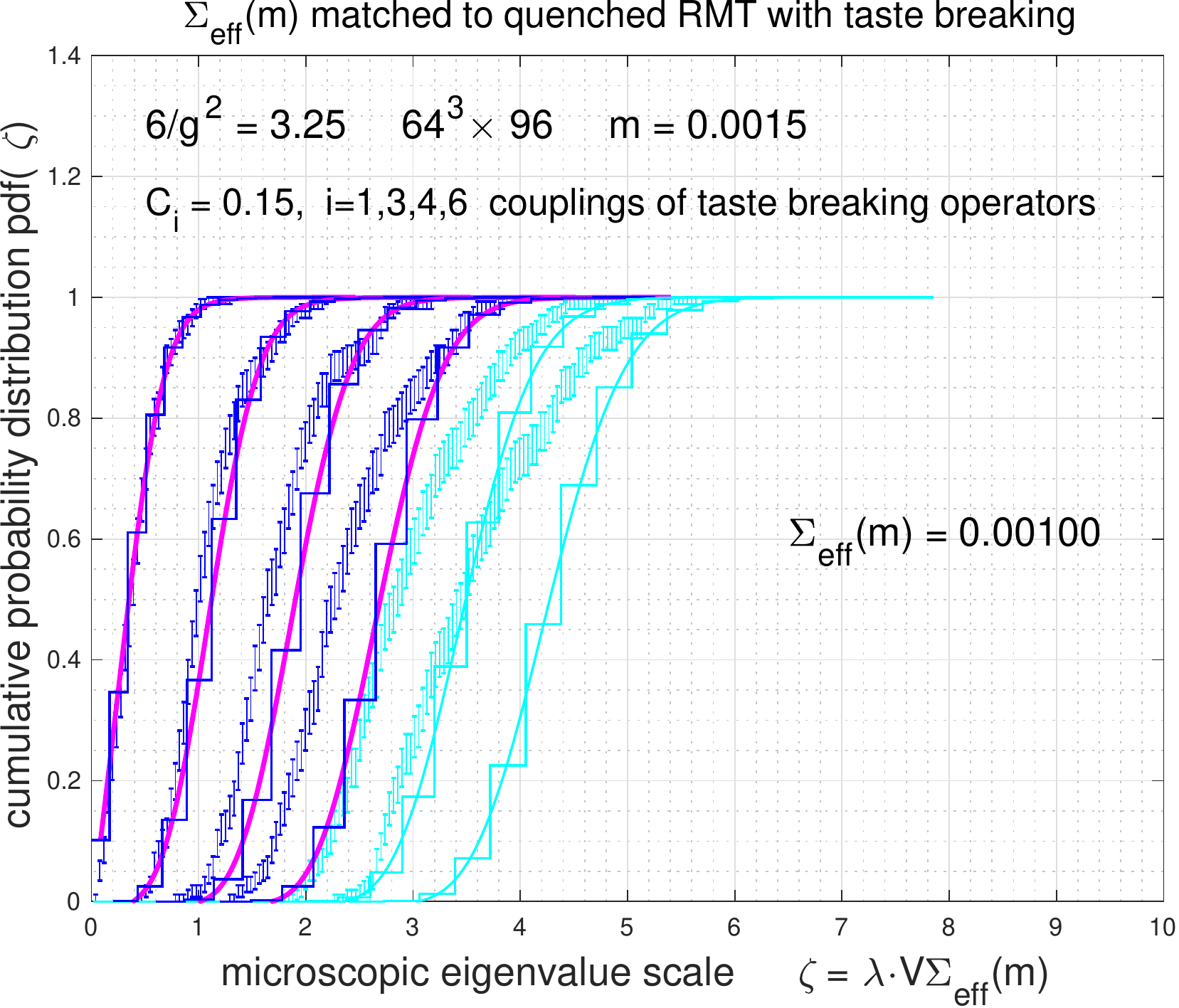}&
				\includegraphics[height=4cm]{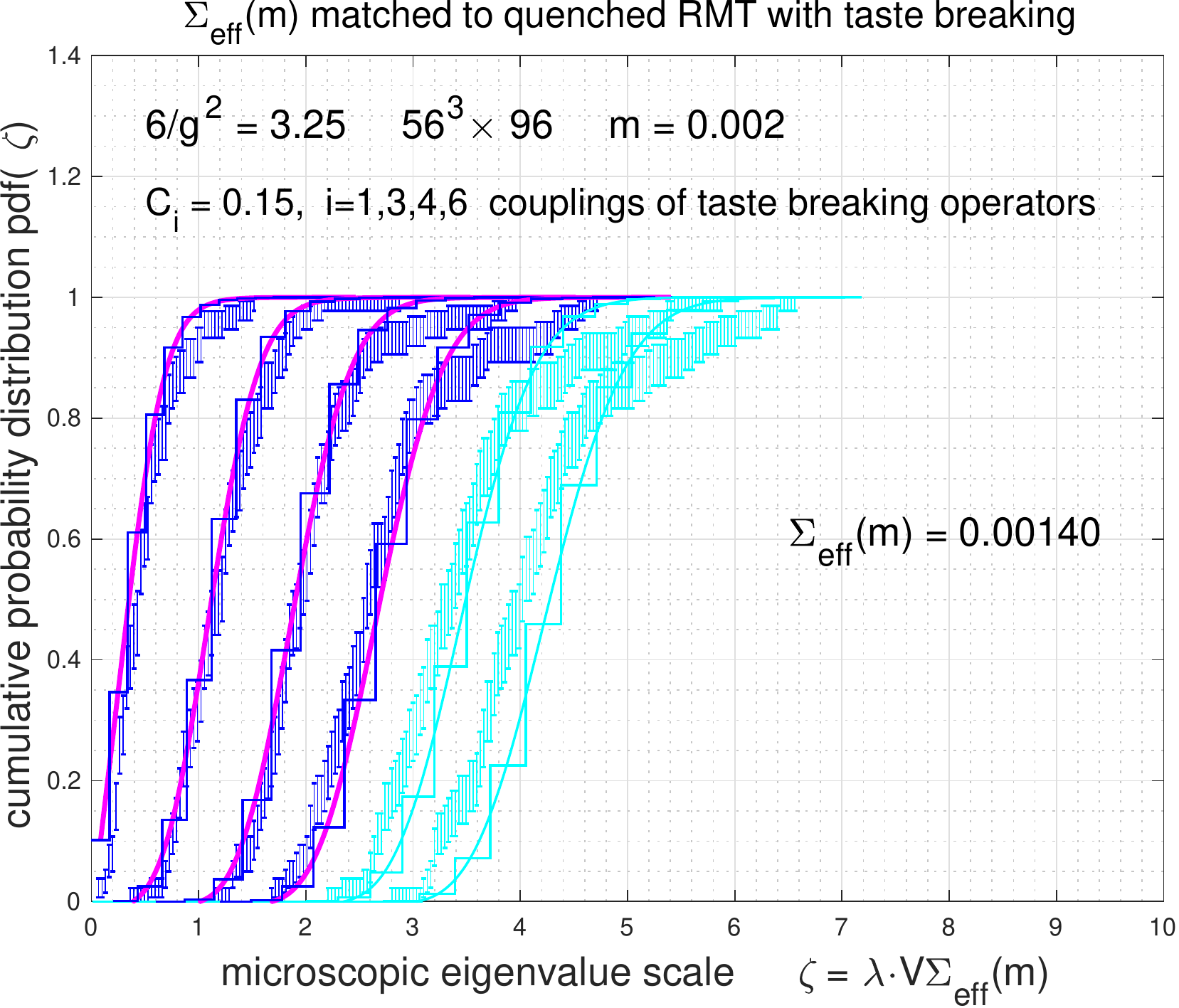}
		    \end{tabular}
		\end{center}	
	    \vskip -0.2in
		\caption{\footnotesize  Microscopic scaling of the lowest Dirac quartet with blue color coding from each of six lattice ensembles is matched  from  lattice configurations with dominantly zero topology  to quenched RMT predictions with cumulative histograms with spline fit in red color to the blue staircase from the RMT simulations. Fermion masses at $m=0.0010/0.0015/0.0020$ were chosen at two lattice spacings with $\beta=3.20/3.25$.  Higher eigenvalues are shown in cyan color with somewhat mismatched cumulative RMT histograms in the same cyan color indicating crossover from the quenched RMT regime to the mesoscopic range of the spectrum above the Thouless energy. The taste breaking coefficients were adjusted differently at the two gauge couplings, consistent with sextet Goldstone spectra at low fermion masses~\cite{Fodor:2016pls}. The combined statistical and systematic errors of $\Sigma(m)$ at all three masses are estimated to remain below 10\%. }
		\label{fig:rmt2}
		\vskip -0.2in
	\end{figure}

We chose the three lowest fermion masses of p-regime ensembles at the largest available volumes and at two different lattice spacings to match the lowest taste-split quartet of Dirac eigenvalues to quenched RMT predictions. The chiral condensate $\Sigma(m)$ of dynamical p-regime simulations is determined at each fermion mass $m$ from the matching microscopic spectrum of the quenched RMT theory, similar to~\cite{Bernardoni:2010nf} and shown in Fig.~\ref{fig:rmt2}. Extrapolations to the chiral limit of the fitted p-regime condensate $\Sigma(m)$ will be compared with direct determination of $\Sigma(0)$ from dynamical lattice ensembles in the $\epsilon$-regime as tested at the extremely low fermion mass. 
%

%%%%%%%%%%%%%%%%%%%%%%%%%%%%%%%%%%%%%%%%%%%%%

%\vskip -0.2in
   {\bf\noindent 5.3 ~~Dynamical RMT analysis with pions in the $\mathbf {\epsilon}$-regime and the dilaton in the p-regime} 
   
\begin{figure}[h!]
	\begin{center}
		\begin{tabular}{cc}
			\includegraphics[height=5cm]{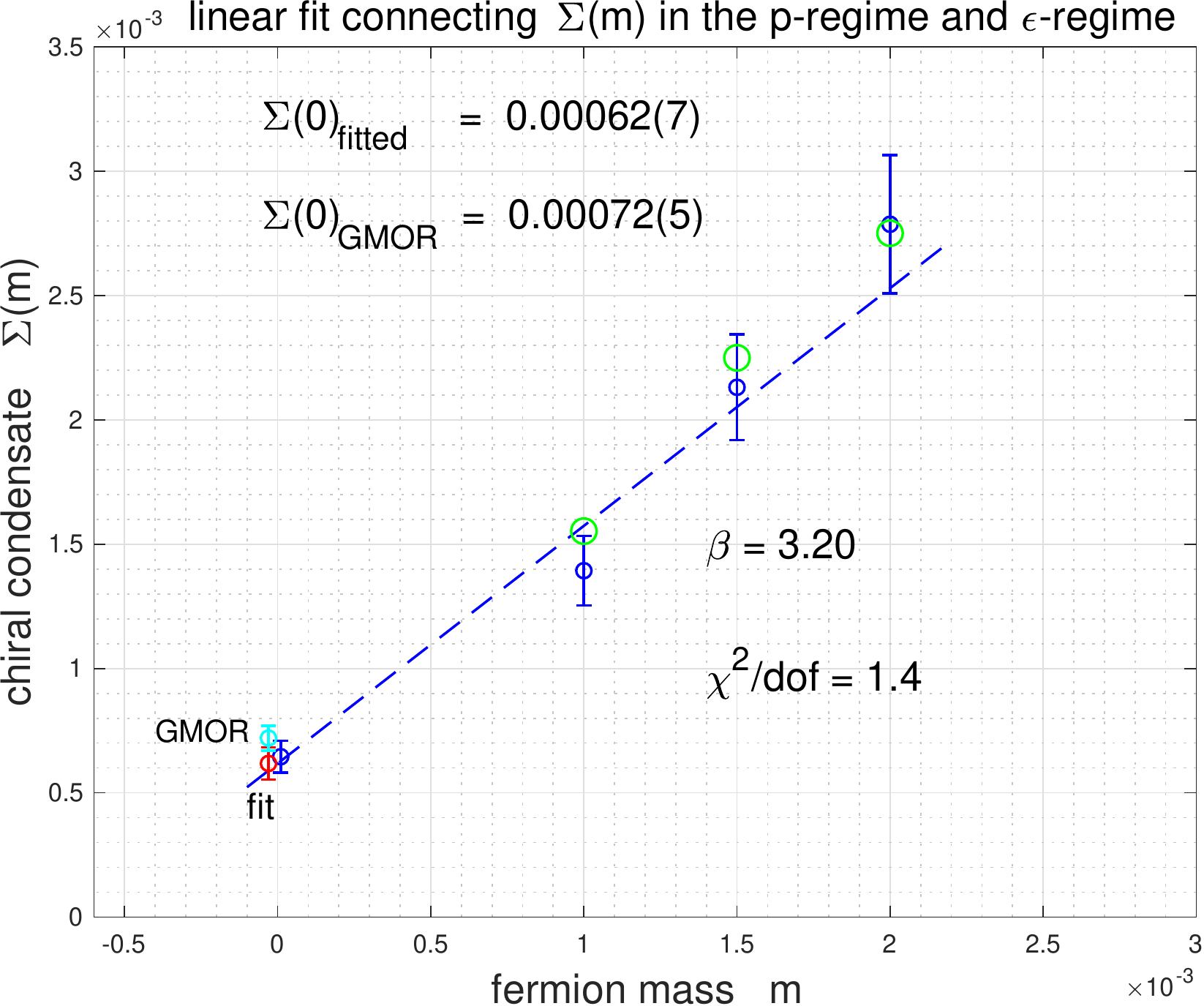}&
			\includegraphics[height=5cm]{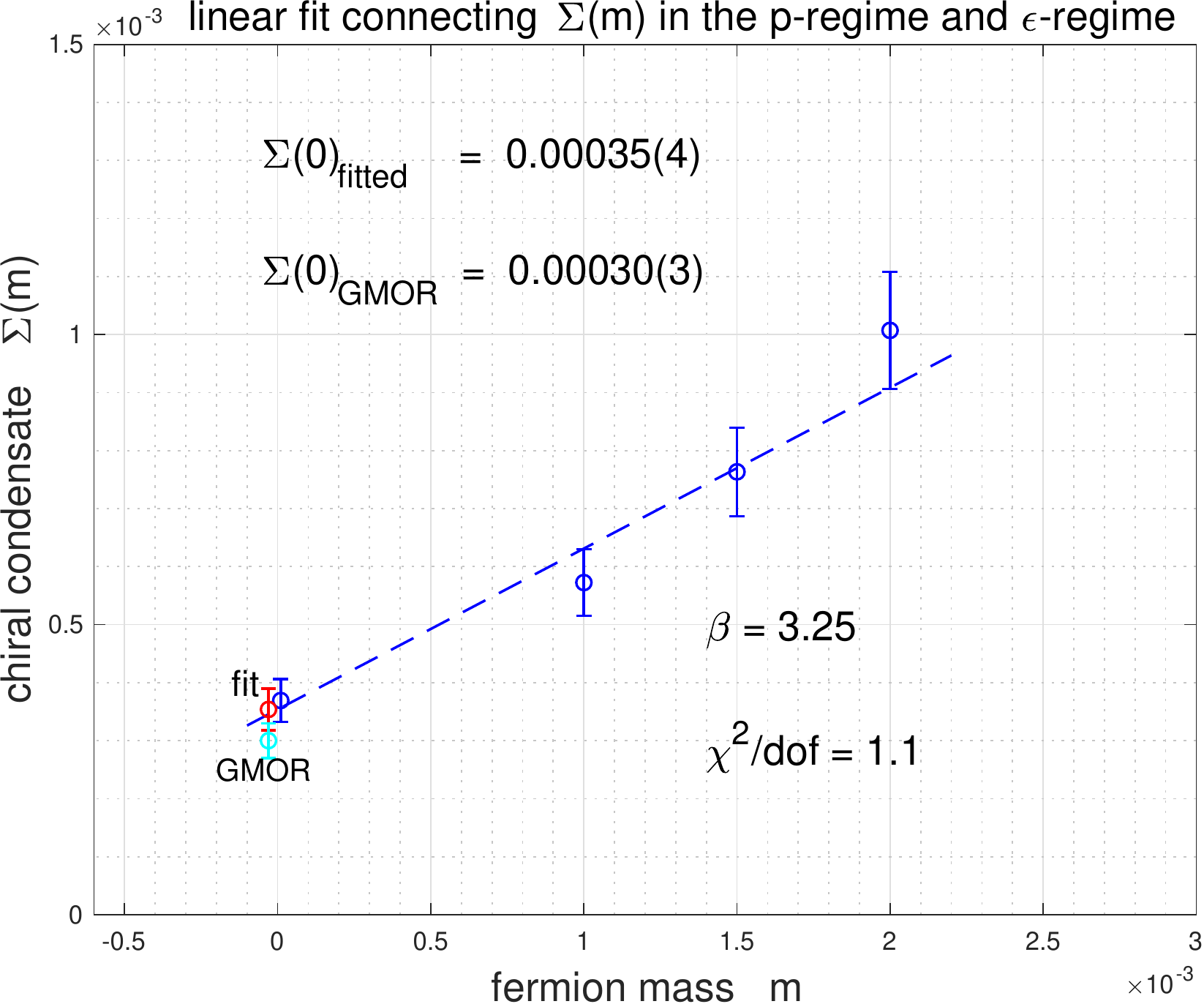}
		\end{tabular}
	\end{center}	
	\vskip -0.1in
	\caption{\footnotesize  The four points where $\Sigma(m)$ was determined in this section is shown in blue color with finite volume corrections applied to $\Sigma$ and with the 10\% error estimate described above. The predicted GMOR values from the p-regime analysis of $V_\sigma$ are also shown. The open green circles on the left panel are the scaled up versions of $\Sigma(0)$ with the $F_d(m)/f_d$ scale factor to the p-regime $m$ values from the dilaton analysis.
	} 
	\label{fig:sigma}
\end{figure}

   Two $\epsilon$-regime ensembles were developed and analyzed here on $64^4$ lattices at a very small $m=0.000010$ fermion mass and at two gauge couplings. 
	\begin{figure}[h!]
		\begin{center}
			\begin{tabular}{cc}
				\includegraphics[height=5cm]{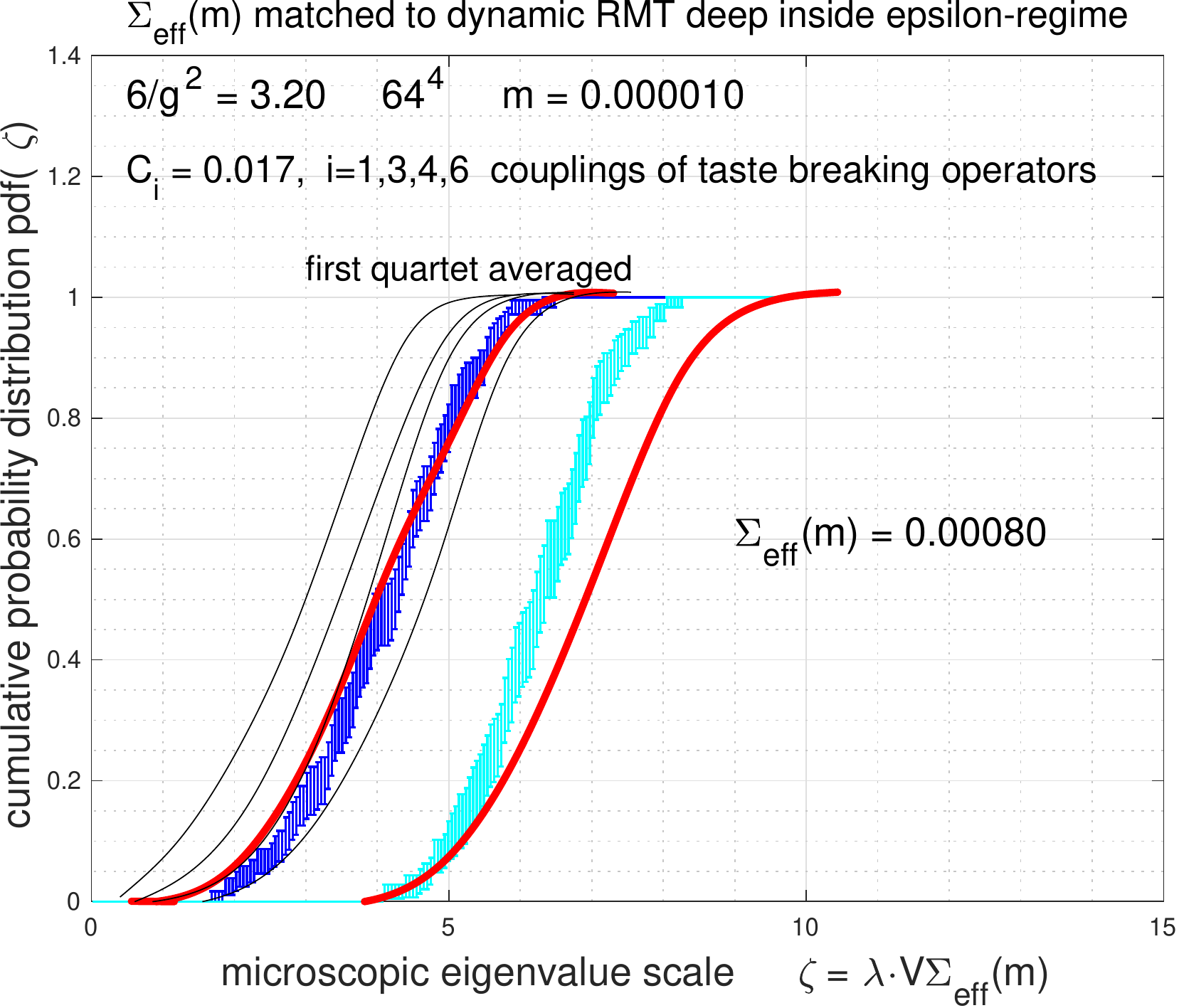}&
				\includegraphics[height=5cm]{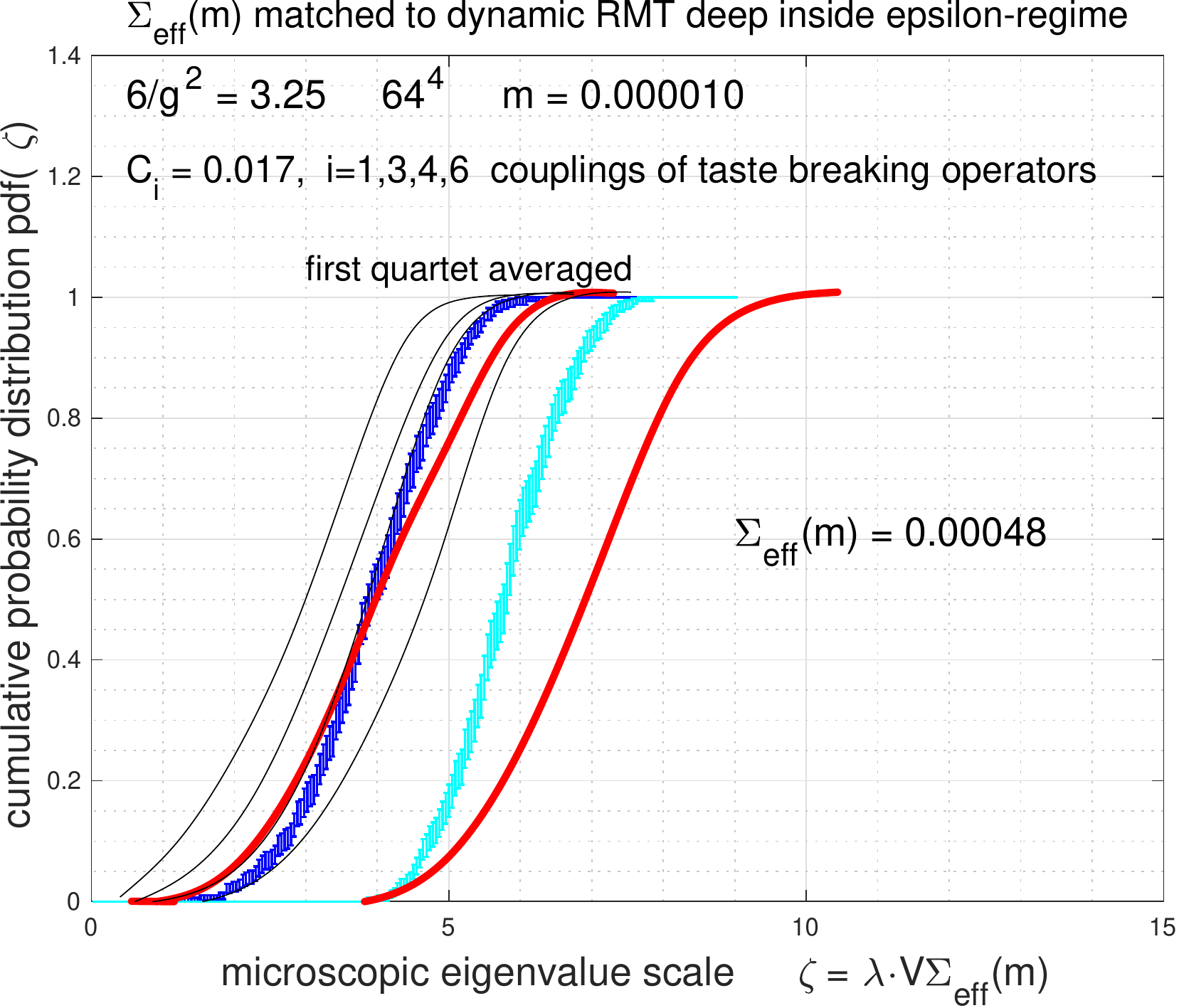}
			\end{tabular}
		\end{center}	
	\vskip -0.2in
		\caption{\footnotesize  Blue color code shows the averaged lowest quartets of the two ensembles  and matched to the RMT prediction plotted with red line before taste-splitting. The analysis is at zero topological charge. Thin black lines show the taste splitting of the dynamical RMT prediction with the fermion determinant included.  The $C_i$ coefficients are significantly lower than in the quenched p-regime analysis, as hinted by the collapse of the sextet Goldstone spectrum toward the chiral limit. The splittings of the individual quartet eigenvalues do not agree well without averaging, worse at $\beta=3.25$.   The second quartet in cyan color is shown for both cases indicating crossover to the mesoscopic regime. 
			The combined statistical and systematic errors of $\Sigma(m)$ at $m=0.000010$ are estimated to remain below 10\% in both cases. 
		} 
		\label{fig:rmt3}
		\vskip -0.2in
	\end{figure}
To match the Dirac spectra to RMT predictions, the dynamical RMT simulations include now the fermion determinants using the re-weighting method which has limited accuracy in predicting the taste-split Dirac spectra. 
We are developing a new HMC based RMT code to improve the stability of the predictions. The averages of four taste-split eigenvalues of the lowest quartet are matched to RMT predictions as shown in Fig.~\ref{fig:rmt3}. The taste splittings of the eigenvalues from the RMT predictions are also shown. 
%\vskip -0.3in
%
Fig.~\ref{fig:sigma} shows the combined results on the chiral condensate from the p-regime and $\epsilon$-regime. 
The three points of the p-regime and the single point of the $\epsilon$-regime can be connected with linear fits at both gauge couplings and warranting explanation  (NLO loop corrections to $\Sigma(m)$  remain untested in this analysis).
%%%%%%%%%%%%%%%%%%%%%%%%%%%%%%%%%%%%%%%%%%%%%
The results show that $\Sigma(m=0.000010)$ is practically in the chiral limit. Agreement of $\Sigma(0)$  is consistent with the GMOR relation from the $V_\sigma$ dilaton potential at both couplings.
  On the left panel of Fig.~\ref{fig:sigma} consistency is shown with the expected dilaton EFT based scaling of the  $\Sigma(m)$ condensate from the p-regime scale factor $F_d(m)/f_d$.
The GMOR prediction for $\Sigma_d(0)$ from the p-regime analysis of the $V_d$ dilaton potential is considerably lower than $\Sigma(0)$ from the RMT analysis. This first look into the $\epsilon$-regime of the dilaton EFT  identified  a problem concerning the $V_d$ potential. Further systematics and statistics is needed for definitive resolution of this problem.

    \noindent{\bf\large 6. ~Brief conclusions:}
    For the first time,  dilaton RMT analysis was applied to a near-conformal gauge theory in the $\epsilon$-regime, reporting results on the chiral condensate $\Sigma(m)$ of the sextet model.
    We demonstrated the sensitivity of the RMT analysis to p-regime predictions for fundamental dilaton EFT parameters from mutually exclusive $f_\pi$ values of two well-motivated dilaton potentials.
    Further work is  ongoing for  direct RMT based determination of $f_\pi$ and for improving  the systematics and the statistical analysis.
    We are also investigating in the sextet model the recently proposed interesting idea to interpolate between the two dilaton potentials $V_\sigma$ and $V_d$ in the p-regime~\cite{Appelquist:2019lgk}. 
%    
%

	%
	
	%%%%%%%%%%%%%%%%%%%%%%%%%%%%%%%%%%%%%%%%%%%%%
	\vskip 0.1in
	\noindent{\bf\large Acknowledgments:}
    Julius Kuti is grateful to James Osborn for extended discussions on the RMT implementation of taste symmetry breaking and for comparisons of computer code output of the related quenched RMT simulations. JK also thanks John McGreevy and Maarten Golterman for discussions. We acknowledge support by the DOE under grant DE-SC0009919, by the NSF under grant 1620845,  and by the Deutsche Forschungsgemeinschaft grant SFB-TR 55. Computational resources were provided by the DOE INCITE program on the ALCF BG/Q platform, by USQCD on gpu platform at Fermilab and BNL, by the University of Wuppertal, and by the Juelich Supercomputing Center. 
    
    %%%%%%%%%%%%%%%%%%%%%%%%%%%%%%%%%%%%%%%%%%%%%

	\begin{spacing}{0.7}
	\bibliographystyle{JHEP}
	\bibliography{jkPoS2019}	

\providecommand{\href}[2]{#2}\begingroup\raggedright\begin{thebibliography}{10}

\bibitem{Fodor:2019vmw}
Z.~Fodor, K.~Holland, J.~Kuti and C.~H. Wong, \emph{{Tantalizing dilaton tests
  from a near-conformal EFT}},
  \href{https://doi.org/10.22323/1.334.0196}{\emph{PoS} {\bfseries LATTICE2018}
  (2019) 196} [\href{https://arxiv.org/abs/1901.06324}{{\ttfamily
  1901.06324}}].

\bibitem{Fodor:2017nlp}
Z.~Fodor, K.~Holland, J.~Kuti, D.~Nogradi and C.~H. Wong, \emph{{The
  twelve-flavor $\beta$-function and dilaton tests of the sextet scalar}},
  \href{https://doi.org/10.1051/epjconf/201817508015}{\emph{EPJ Web Conf.}
  {\bfseries 175} (2018) 08015}
  [\href{https://arxiv.org/abs/1712.08594}{{\ttfamily 1712.08594}}].

\bibitem{Hong:2004td}
D.~K. Hong, S.~D.~H. Hsu and F.~Sannino, \emph{{Composite Higgs from higher
  representations}},
  \href{https://doi.org/10.1016/j.physletb.2004.07.007}{\emph{Phys. Lett.}
  {\bfseries B597} (2004) 89}
  [\href{https://arxiv.org/abs/hep-ph/0406200}{{\ttfamily hep-ph/0406200}}].

\bibitem{Dietrich:2005jn}
D.~D. Dietrich, F.~Sannino and K.~Tuominen, \emph{{Light composite Higgs from
  higher representations versus electroweak precision measurements: Predictions
  for CERN LHC}}, \href{https://doi.org/10.1103/PhysRevD.72.055001}{\emph{Phys.
  Rev.} {\bfseries D72} (2005) 055001}
  [\href{https://arxiv.org/abs/hep-ph/0505059}{{\ttfamily hep-ph/0505059}}].

\bibitem{Fodor:2012ty}
Z.~Fodor, K.~Holland, J.~Kuti, D.~Nogradi, C.~Schroeder and C.~H. Wong,
  \emph{{Can the nearly conformal sextet gauge model hide the Higgs
  impostor?}},
  \href{https://doi.org/10.1016/j.physletb.2012.10.079}{\emph{Phys. Lett.}
  {\bfseries B718} (2012) 657}
  [\href{https://arxiv.org/abs/1209.0391}{{\ttfamily 1209.0391}}].

\bibitem{Fodor:2014pqa}
Z.~Fodor, K.~Holland, J.~Kuti, D.~Nogradi and C.~H. Wong, \emph{{Can a light
  Higgs impostor hide in composite gauge models?}}, {\emph{PoS} {\bfseries
  LATTICE2013} (2014) 062}.

\bibitem{Fodor:2016pls}
Z.~Fodor, K.~Holland, J.~Kuti, S.~Mondal, D.~Nogradi and C.~H. Wong,
  \emph{{Status of a minimal composite Higgs theory}},
  \href{https://doi.org/10.22323/1.251.0219}{\emph{PoS} {\bfseries LATTICE2015}
  (2016) 219} [\href{https://arxiv.org/abs/1605.08750}{{\ttfamily
  1605.08750}}].

\bibitem{Golterman:2016lsd}
M.~Golterman and Y.~Shamir, \emph{{Low-energy effective action for pions and a
  dilatonic meson}},
  \href{https://doi.org/10.1103/PhysRevD.94.054502}{\emph{Phys. Rev.}
  {\bfseries D94} (2016) 054502}
  [\href{https://arxiv.org/abs/1603.04575}{{\ttfamily 1603.04575}}].

\bibitem{Golterman:2016cdd}
M.~Golterman and Y.~Shamir, \emph{{Effective pion mass term and the trace
  anomaly}}, \href{https://doi.org/10.1103/PhysRevD.95.016003}{\emph{Phys.
  Rev.} {\bfseries D95} (2017) 016003}
  [\href{https://arxiv.org/abs/1611.04275}{{\ttfamily 1611.04275}}].

\bibitem{Appelquist:2017wcg}
T.~Appelquist, J.~Ingoldby and M.~Piai, \emph{{Dilaton EFT Framework For
  Lattice Data}}, \href{https://doi.org/10.1007/JHEP07(2017)035}{\emph{JHEP}
  {\bfseries 07} (2017) 035}
  [\href{https://arxiv.org/abs/1702.04410}{{\ttfamily 1702.04410}}].

\bibitem{Appelquist:2017vyy}
T.~Appelquist, J.~Ingoldby and M.~Piai, \emph{{Analysis of a Dilaton EFT for
  Lattice Data}}, \href{https://doi.org/10.1007/JHEP03(2018)039}{\emph{JHEP}
  {\bfseries 03} (2018) 039}
  [\href{https://arxiv.org/abs/1711.00067}{{\ttfamily 1711.00067}}].

\bibitem{Golterman:2018mfm}
M.~Golterman and Y.~Shamir, \emph{{The large-mass regime of the dilaton-pion
  low-energy effective theory}},
  \href{https://doi.org/10.1103/PhysRevD.98.056025}{\emph{Phys. Rev.}
  {\bfseries D98} (2018) 056025}
  [\href{https://arxiv.org/abs/1805.00198}{{\ttfamily 1805.00198}}].

\bibitem{Golterman:2018bpc}
M.~Golterman and Y.~Shamir, \emph{{The large-mass regime of confining but
  nearly conformal gauge theories}},  in \emph{{Lattice 2018}}, 2018,
  \href{https://arxiv.org/abs/1810.05353}{{\ttfamily 1810.05353}}.

\bibitem{Appelquist:2019lgk}
T.~Appelquist, J.~Ingoldby and M.~Piai, \emph{{The Dilaton Potential and
  Lattice Data}},  \href{https://arxiv.org/abs/1908.00895}{{\ttfamily
  1908.00895}}.

\bibitem{Fodor:2015zna}
Z.~Fodor, K.~Holland, J.~Kuti, S.~Mondal, D.~Nogradi and C.~H. Wong, \emph{{The
  running coupling of the minimal sextet composite Higgs model}},
  \href{https://doi.org/10.1007/JHEP09(2015)039}{\emph{JHEP} {\bfseries 09}
  (2015) 039} [\href{https://arxiv.org/abs/1506.06599}{{\ttfamily
  1506.06599}}].

\bibitem{Fodor:2019ypi}
Z.~Fodor, K.~Holland, J.~Kuti, D.~Nogradi and C.~H. Wong, \emph{{Case studies
  of near-conformal $\beta$-functions}},  in \emph{{37th International
  Symposium on Lattice Field Theory (Lattice 2019) Wuhan, Hubei, China, June
  16-22, 2019}}, 2019, \href{https://arxiv.org/abs/1912.07653}{{\ttfamily
  1912.07653}}.

\bibitem{Fodor:2016hke}
Z.~Fodor, K.~Holland, J.~Kuti, S.~Mondal, D.~Nogradi and C.~H. Wong, \emph{{New
  approach to the Dirac spectral density in lattice gauge theory
  applications}}, {\emph{PoS} {\bfseries LATTICE2015} (2016) 310}
  [\href{https://arxiv.org/abs/1605.08091}{{\ttfamily 1605.08091}}].

\bibitem{Golterman:2019htd}
M.~Golterman and Y.~Shamir, \emph{{Fits of $SU(3)$ $N_f=8$ data to dilaton-pion
  effective field theory}},  in \emph{{37th International Symposium on Lattice
  Field Theory (Lattice 2019) Wuhan, Hubei, China, June 16-22, 2019}}, 2019,
  \href{https://arxiv.org/abs/1910.10331}{{\ttfamily 1910.10331}}.

\bibitem{Brown:2019ipr}
T.~V. Brown, M.~Golterman, S.~Krøjer, Y.~Shamir and K.~Splittorff, \emph{{The
  $\epsilon$-regime of dilaton chiral perturbation theory}},
  \href{https://doi.org/10.1103/PhysRevD.100.114515}{\emph{Phys. Rev.}
  {\bfseries D100} (2019) 114515}
  [\href{https://arxiv.org/abs/1909.10796}{{\ttfamily 1909.10796}}].

\bibitem{Vecchi:2010jz}
L.~Vecchi, \emph{{The Conformal Window of deformed CFT's in the planar limit}},
  \href{https://doi.org/10.1103/PhysRevD.82.045013}{\emph{Phys. Rev.}
  {\bfseries D82} (2010) 045013}
  [\href{https://arxiv.org/abs/1004.2063}{{\ttfamily 1004.2063}}].

\bibitem{Gorbenko:2018ncu}
V.~Gorbenko, S.~Rychkov and B.~Zan, \emph{{Walking, Weak first-order
  transitions, and Complex CFTs}},
  \href{https://doi.org/10.1007/JHEP10(2018)108}{\emph{JHEP} {\bfseries 10}
  (2018) 108} [\href{https://arxiv.org/abs/1807.11512}{{\ttfamily
  1807.11512}}].

\bibitem{Gorbenko:2018dtm}
V.~Gorbenko, S.~Rychkov and B.~Zan, \emph{{Walking, Weak first-order
  transitions, and Complex CFTs II. Two-dimensional Potts model at $Q>4$}},
  \href{https://doi.org/10.21468/SciPostPhys.5.5.050}{\emph{SciPost Phys.}
  {\bfseries 5} (2018) 050} [\href{https://arxiv.org/abs/1808.04380}{{\ttfamily
  1808.04380}}].

\bibitem{Banks:1981nn}
T.~Banks and A.~Zaks, \emph{{On the Phase Structure of Vector-Like Gauge
  Theories with Massless Fermions}},
  \href{https://doi.org/10.1016/0550-3213(82)90035-9}{\emph{Nucl. Phys.}
  {\bfseries B196} (1982) 189}.

\bibitem{Kaplan:2009kr}
D.~B. Kaplan, J.-W. Lee, D.~T. Son and M.~A. Stephanov, \emph{{Conformality
  Lost}}, \href{https://doi.org/10.1103/PhysRevD.80.125005}{\emph{Phys. Rev.}
  {\bfseries D80} (2009) 125005}
  [\href{https://arxiv.org/abs/0905.4752}{{\ttfamily 0905.4752}}].

\bibitem{Gies:2005as}
H.~Gies and J.~Jaeckel, \emph{{Chiral phase structure of QCD with many
  flavors}}, \href{https://doi.org/10.1140/epjc/s2006-02475-0}{\emph{Eur. Phys.
  J.} {\bfseries C46} (2006) 433}
  [\href{https://arxiv.org/abs/hep-ph/0507171}{{\ttfamily hep-ph/0507171}}].

\bibitem{Baxter:1978}
R.~J. Baxter, \emph{Variational approximations for square lattice models in
  statistical mechanics}, {\emph{J. Stat. Phys.} {\bfseries 19} (1978) 461}.

\bibitem{Nishino:1996}
T.~Nishino and K.~Okunishi, \emph{{Corner Transfer Matrix Renormalization Group
  method}}, {\emph{Journal of the Physical Society of Japan} {\bfseries 65}
  (1996) 891}.

\bibitem{Orlando:2019skh}
D.~Orlando, S.~Reffert and F.~Sannino, \emph{{Near-Conformal Dynamics at Large
  Charge}},  \href{https://arxiv.org/abs/1909.08642}{{\ttfamily 1909.08642}}.

\bibitem{Goldberger:2008zz}
W.~D. Goldberger, B.~Grinstein and W.~Skiba, \emph{{Distinguishing the Higgs
  boson from the dilaton at the Large Hadron Collider}},
  \href{https://doi.org/10.1103/PhysRevLett.100.111802}{\emph{Phys. Rev. Lett.}
  {\bfseries 100} (2008) 111802}
  [\href{https://arxiv.org/abs/0708.1463}{{\ttfamily 0708.1463}}].

\bibitem{Migdal:1982jp}
A.~A. Migdal and M.~A. Shifman, \emph{{Dilaton Effective Lagrangian in
  Gluodynamics}},
  \href{https://doi.org/10.1016/0370-2693(82)90089-2}{\emph{Phys. Lett.}
  {\bfseries 114B} (1982) 445}.

\bibitem{Ellis:1984jv}
J.~R. Ellis and J.~Lanik, \emph{{IS SCALAR GLUONIUM OBSERVABLE?}},
  \href{https://doi.org/10.1016/0370-2693(85)91013-5}{\emph{Phys. Lett.}
  {\bfseries 150B} (1985) 289}.

\bibitem{Bardeen:1985sm}
W.~A. Bardeen, C.~N. Leung and S.~T. Love, \emph{{The Dilaton and Chiral
  Symmetry Breaking}},
  \href{https://doi.org/10.1103/PhysRevLett.56.1230}{\emph{Phys. Rev. Lett.}
  {\bfseries 56} (1986) 1230}.

\bibitem{Leung:1989hw}
C.~N. Leung, S.~T. Love and W.~A. Bardeen, \emph{{Aspects of Dynamical Symmetry
  Breaking in Gauge Field Theories}},
  \href{https://doi.org/10.1016/0550-3213(89)90121-1}{\emph{Nucl. Phys.}
  {\bfseries B323} (1989) 493}.

\bibitem{Donoghue:1991qv}
J.~F. Donoghue and H.~Leutwyler, \emph{{Energy and momentum in chiral
  theories}}, \href{https://doi.org/10.1007/BF01560453}{\emph{Z. Phys.}
  {\bfseries C52} (1991) 343}.

\bibitem{Chacko:2012sy}
Z.~Chacko and R.~K. Mishra, \emph{{Effective Theory of a Light Dilaton}},
  {\emph{Phys. Rev.} {\bfseries D87} (2013) 115006.}

\bibitem{Vecchi:2010gj}
L.~Vecchi, \emph{{Phenomenology of a light scalar: the dilaton}}, {\emph{Phys.
  Rev.} {\bfseries D82} (2010) 076009.}

\bibitem{Matsuzaki:2013eva}
S.~Matsuzaki and K.~Yamawaki, \emph{{Dilaton Chiral Perturbation Theory:
  Determining the Mass and Decay Constant of the Technidilaton on the
  Lattice}}, {\emph{Phys. Rev. Lett.} {\bfseries 113} (2014) 082002.}

\bibitem{Aoki:2016wnc}
{\scshape LatKMI} collaboration, Y.~Aoki et~al., \emph{{Light flavor-singlet
  scalars and walking signals in $N_f=8$ QCD on the lattice}},
  \href{https://doi.org/10.1103/PhysRevD.96.014508}{\emph{Phys. Rev.}
  {\bfseries D96} (2017) 014508}
  [\href{https://arxiv.org/abs/1610.07011}{{\ttfamily 1610.07011}}].

\bibitem{Hansen:2016fri}
M.~Hansen, K.~Langæble and F.~Sannino, \emph{{Extending Chiral Perturbation
  Theory with an Isosinglet Scalar}},
  \href{https://doi.org/10.1103/PhysRevD.95.036005}{\emph{Phys. Rev.}
  {\bfseries D95} (2017) 036005}
  [\href{https://arxiv.org/abs/1610.02904}{{\ttfamily 1610.02904}}].

\bibitem{Cata:2018wzl}
O.~Catà, R.~J. Crewther and L.~C. Tunstall, \emph{{Crawling technicolor}},
  \href{https://arxiv.org/abs/1803.08513}{{\ttfamily 1803.08513}}.

\bibitem{Fodor:2017wsn}
Z.~Fodor, K.~Holland, J.~Kuti, D.~Nogradi and C.~H. Wong, \emph{{Spectroscopy
  of the BSM sextet model}},
  \href{https://doi.org/10.1051/epjconf/201817508014}{\emph{EPJ Web Conf.}
  {\bfseries 175} (2018) 08014}
  [\href{https://arxiv.org/abs/1711.05299}{{\ttfamily 1711.05299}}].

\bibitem{Bernardoni:2010nf}
F.~Bernardoni, P.~Hernandez, N.~Garron, S.~Necco and C.~Pena, \emph{{Probing
  the chiral regime of $N_{f}$= 2 QCD with mixed actions}},
  \href{https://doi.org/10.1103/PhysRevD.83.054503}{\emph{Phys. Rev.}
  {\bfseries D83} (2011) 054503}
  [\href{https://arxiv.org/abs/1008.1870}{{\ttfamily 1008.1870}}].

\bibitem{Osborn:2010eq}
J.~C. Osborn, \emph{{Staggered chiral random matrix theory}},
  \href{https://doi.org/10.1103/PhysRevD.83.034505}{\emph{Phys. Rev.}
  {\bfseries D83} (2011) 034505}
  [\href{https://arxiv.org/abs/1012.4837}{{\ttfamily 1012.4837}}].

\bibitem{Osborn:2012bc}
J.~C. Osborn, \emph{{Chiral random matrix theory for staggered fermions}},
  \href{https://doi.org/10.22323/1.139.0110}{\emph{PoS} {\bfseries LATTICE2011}
  (2011) 110} [\href{https://arxiv.org/abs/1204.5497}{{\ttfamily 1204.5497}}].

\end{thebibliography}\endgroup
	\end{spacing}
	
\end{document}